\documentclass[english,12pt,a4paper]{article}
\usepackage{authblk}
\usepackage[text={17.0cm,23.7cm}]{geometry}
\usepackage{babel}
\usepackage[colorlinks]{hyperref}
\usepackage{cite}
\usepackage{graphicx}
\pdfoutput=1
\usepackage{amsmath}
\usepackage{amssymb}
\usepackage[dvipsnames]{xcolor}
\usepackage{yfonts}
\usepackage{ulem}
\usepackage[font=small,labelfont=bf]{caption}

\newcommand\myshade{80}
\colorlet{mylinkcolor}{ForestGreen}
\colorlet{mycitecolor}{Red}
\colorlet{myurlcolor}{violet}

\hypersetup{
  linkcolor  = mylinkcolor!\myshade!black,
  citecolor  = mycitecolor!\myshade!black,
  urlcolor   = myurlcolor!\myshade!black,
  colorlinks = true
}

\newcommand{\mdm}{m_\text{DM}}
\newcommand{\GeVcm}{\text{GeV}/\text{cm}^3}
\newcommand{\vecvobs}{\vec{v}_\text{obs}}

\newcommand{\ER}{E_R}
\newcommand{\dER}{\text{d}E_R}
\newcommand{\mNi}{m_{N_i}}
\newcommand{\rholoc}{\rho_\text{loc}}

\newcommand{\Nobs}{N_\mathrm{obs}}
\newcommand{\Nsig}{N_\mathrm{sig}}
\newcommand{\NBG}{N_\mathrm{BG}}

\begin{document}
\title{\vspace{-2cm}
{\normalsize
\flushright TUM-HEP 1147/18,\,KIAS-P18061\\}
\vspace{0.6cm}
\textbf{Bracketing the impact of astrophysical uncertainties on local dark matter searches}\\[8mm]}

\author[1,2]{Alejandro Ibarra}
\author[3]{Bradley J. Kavanagh}
\author[1]{Andreas Rappelt}
\affil[1]{\normalsize\textit{Physik-Department T30d, Technische Universit\"at M\"unchen, James-Franck-Stra\ss{}e, 85748 Garching, Germany}}
\affil[2]{\normalsize\textit{School of Physics, Korea Institute for Advanced Study, Seoul 02455, South Korea}}
\affil[3]{\normalsize\textit{GRAPPA, Institute of Physics, University of Amsterdam, 1098 XH Amsterdam, The Netherlands}}

\date{}

\maketitle
\begin{abstract}
	The theoretical interpretation of dark matter (DM) experiments is hindered by uncertainties on the dark matter density and velocity distribution inside the Solar System. In order to quantify those uncertainties, we present a parameter that characterizes the deviation of the true velocity distribution from the standard Maxwell-Boltzmann form, and we then determine for different values of this parameter the most aggressive and most conservative limits on the dark matter scattering cross section with nuclei; uncertainties in the local dark matter density can be accounted for trivially. This allows us to bracket, in a model independent way, the impact of astrophysical uncertainties on limits from direct detection experiments and/or neutrino telescopes. We find that current limits assuming the Standard Halo Model are at most a factor of $\sim 2$ weaker than the most aggressive possible constraints. In addition, combining neutrino telescope and direct detection constraints (in a statistically meaningful way), we show that limits on DM in the mass range $\sim 10 - 1000$ GeV cannot be weakened by more than around a factor of 10, for all possible velocity distributions. We finally demonstrate that our approach can also be employed in the event of a DM discovery, allowing us to avoid bias in the reconstruction of the DM properties.
\end{abstract}

\section{Introduction}

The identity and nature of dark matter (DM) is one of the biggest puzzles in modern cosmology \cite{Bertone:2004pz} and a great deal of experimental effort is dedicated to solving it \cite{Bauer:2013ihz}. The hunt for particle DM includes (but is not limited to) collider searches for DM production \cite{Kahlhoefer:2017dnp}, indirect searches for DM annihilation products \cite{Gaskins:2016cha} and direct searches for DM scattering off Standard Model particles in the lab \cite{Undagoitia:2015gya}. Because we know so little about DM itself, it is not always clear what a DM signal will look like, meaning that searches can suffer from large theoretical uncertainties (in addition to non-trivial experimental uncertainties). Understanding and quantifying these uncertainties is crucial for interpreting exclusion limits from DM searches and for building a consistent picture in the event of a discovery.

In this work, we focus on astrophysical uncertainties arising in direct detection experiments (lab-based searches for DM-nucleus or DM-electron scattering \cite{Goodman:1984dc,Drukier:1986tm}) and Solar capture (which can produce an observable neutrino signal from annihilation of captured DM  \cite{Press:1985ug,Silk:1985ax}). The overall normalization of DM signals from direct detection and Solar capture depends on the local DM density in the Solar neighbourhood, for which a fixed reference value of $\rholoc = 0.3 \;\GeVcm$  is typically assumed \cite{Green:2011bv}. The precise value of the dark matter density inside the Solar System is, however, unknown. Astrophysical estimates of $\rholoc$ suggest a factor of a few uncertainty \cite{Read:2014qva}: global estimates from mass modelling of the Milky Way point towards $0.2\,-\,0.4\;\GeVcm$ \cite{Salucci:2010qr,Catena:2009mf,Weber2010,Iocco:2011jz,McMillan2011,Pato:2015dua,Huang2016,McMillan2017} while local measurements from stellar kinematics (at distances $\lesssim 1$ kpc from the Sun) find larger values, reaching up to 0.85 $\GeVcm$  \cite{MoniBidin2012,Bovy:2012tw,Garbari2012,Smith2012,Zhang2013,Bovy:2013raa,Bienayme:2014kva,McKee:2015hwa,Xia:2015agz,Sivertsson:2017rkp}. Uncertainties on $\rholoc$ can often be accounted for trivially, by rescaling the signal normalization appropriately. 

Direct detection and Solar capture signals also depend on the local velocity distribution $f(\vec{v})$ of dark matter\footnote{Sometimes the distinction is made between the \textit{velocity} $\vec{v}$ of DM particles and their \textit{speed} $v \equiv |\vec{v}|$. Sensitivity to the velocity vector $\vec{v}$ requires directionally-sensitive detectors \cite{Mayet:2016zxu}, which we do not discuss in detail here. We will therefore use the terms \textit{velocity} and \textit{speed} interchangeably.} : faster moving particles typically produce more energetic recoils \cite{Ferrer:2015bta} while slow moving particles are more easily captured by the Sun \cite{Choi:2013eda,Danninger:2014xza}. A common benchmark is the Standard Halo Model (SHM), in which the DM follows a Maxwell-Boltzmann (MB) velocity distribution $f_\mathrm{MB}(\vec{v})$, valid for a spherically-symmetric, equilibrium DM halo with a $\rho \propto r^{-2}$ density profile (see e.g.~Appendix A of Ref.~\cite{Peter-Thesis}). Deviations from this simple model of the DM halo are expected in the Milky Way \cite{Bhattacharjee:2012xm,Fornasa:2013iaa,Bozorgnia:2013pua,Mandal:2018efq} and while the parameters associated with the SHM (such as the Sun's velocity \cite{Bovy:2012ba}, the local circular velocity \cite{Koposov:2009hn} and the Galactic escape velocity \cite{Piffl:2013mla}) can be estimated observationally, they carry with them an associated uncertainty \cite{Green:2011bv,Benito:2016kyp,Green:2017odb}.  In addition, numerical simulations have suggested the possibility of non-Maxwellian structure in the DM velocity distribution \cite{Vogelsberger:2008qb,Kuhlen:2009vh,Read:2009iv,Read:2009jy,Mao:2012hf,Kuhlen:2012fz,Kuhlen:2013tra,Butsky:2015pya}. While some state-of-the-art hydrodynamical simulations find DM distributions consistent with the Maxwell-Boltzmann form \cite{Bozorgnia:2016ogo,Bozorgnia:2017brl}, it is still possible that ultra-local substructures such as streams may also contribute \cite{Freese:2003na,Purcell:2012sh}.

There have been a range of studies assessing the impact of these uncertainties \cite{Fairbairn:2012zs} and proposing methods to overcome them \cite{Peter:2009ak,Cremonesi:2013bma}. These approaches aim to interpret DM searches in as general a way as possible, being agnostic about the shape of $f(\vec{v})$. For a long time, the standard halo-independent approach initiated by Fox et al.~\cite{Fox:2010bz,Fox:2010bu,McCabe:2011sr,Frandsen:2011gi,HerreroGarcia:2011aa,Gondolo:2012rs,DelNobile:2013cta,Cherry:2014wia,Kahlhoefer:2016eds} could not be used to give concrete statistical conclusions, though more recent work has begun to address this issue (see e.g.~Refs\cite{Gelmini:2015voa,Gelmini:2016pei,Gondolo:2017jro,Gelmini:2017aqe}). Certain astrophysics-independent approaches have been introduced which do have a robust statistical meaning, though these are typically computationally expensive (e.g.~Refs.~\cite{Peter:2011eu,Kavanagh:2012nr,Kavanagh:2013wba,Kavanagh:2013eya,Kavanagh:2014rya}) or have not been applied to neutrino telescope data (e.g.~Refs.~\cite{Feldstein:2014gza,Feldstein:2014ufa}). 

Reference~\cite{Ibarra:2017mzt} (hereafter referred to as IR17) developed a new fully halo-independent approach to comparing direct detection and neutrino telescope experiments. Noting that the nuclear scattering rate (in the case of direct detection) and solar capture rate (relevant for neutrino telescopes) are both linear in the DM velocity distribution $f(\vec{v})$, IR17 showed that the techniques of linear programming can be applied to optimize the signal in one experiment, subject to the constraints imposed by another experiment. In simple terms, the velocity distribution is written as a large number of velocity streams whose weights are optimized to maximize or minimize the number of signal events, subject to a set of constraints. In the end, the optimized solution is described by only a small number of streams, at most as many as there are constraints in the optimization problem.

In this work, we refine the approach of IR17 and extend its application to understanding how arbitrary distortions in the DM velocity distribution affect limits from DM search experiments. In order to do this, we first develop a technique to assign robust statistical meaning to halo-independent limits, now using techniques from quadratic programming. We then explore how the addition of DM streams to the (typically assumed) smooth Maxwell-Boltzmann velocity distribution affects upper limits from current experiments. In addition, we examine how combined limits from multiple experiments vary as the MB distribution is smoothly perturbed, allowing us to determine how the limits vary as a function of `closeness' to the Standard Halo Model (SHM) (see~Ref.~\cite{Fowlie:2017ufs} for a related analysis using Bayesian statistics). Finally, equipped with a robust statistical interpretation of our method, we explore how astrophysical uncertainties will affect the reconstruction of DM properties from a future detection. Ultimately, this approach allows us to bracket -- in an efficient and fully halo-independent way -- the impact of astrophysical uncertainties on results from DM searches.

We begin in Sec.~\ref{sec:probes_of_dark_matter} with a review of dark matter signals from direct detection and Solar capture. We then discuss in Sec.~\ref{sec:optimization} the optimization methods we use, first introduced in IR17 and extended here. In Sec.~\ref{sec:Applications}, we present a number of applications of these methods, including the calculation of both robust exclusion limits and parameter reconstruction with future experiments. Finally, in Sec.~\ref{sec:Conclusions} we discuss the implications of our approach and present our conclusions.

\section{Probes of dark matter inside the Solar System}
\label{sec:probes_of_dark_matter}

Several methods have been proposed to probe the particle nature of the dark matter population inside the Solar System, under the assumption that the DM interacts with nuclei. In order to interpret such searches, we must make some assumptions about the DM distribution. We postulate that the Solar System is embedded in a stationary and spatially homogeneous dark matter halo, with mass density $\rho_{\rm loc}$ and velocity distribution relative to the Solar frame $f(\vec v)$,  normalized such that
\begin{equation}
\int_{v~\leq v_{\text{max}}}\;\text{d}v^{3}\;f(\vec{v})=1\,.
\label{eq:VDNormalization}
\end{equation}
Here,  $v\equiv |\vec v|$ and $v_{\text{max}} = v_\mathrm{esc} + v_\odot$ is the maximal velocity of a dark matter particle that is gravitationally bound to the galaxy (expressed in the Solar frame), which can be calculated from the galactic escape velocity $v_\mathrm{esc}$, taken to be in the range  499-608 ${\rm km}/{\rm s}$~\cite{Smith:2006ym,Piffl:2013mla}, and the local velocity of the Sun with respect to the halo $v_\odot \simeq 244$ km/s~\cite{Xue:2008se,McMillan:2009yr,Bovy:2009dr}. 

The assumptions given so far are relatively weak. We now turn to a much stronger assumption which is often made. Under the Standard Halo Model (SHM), the velocity distribution takes the form of a Maxwell-Boltzmann (MB) distribution:
\begin{align}
\label{eq:f_MB}
f_\mathrm{MB} ( \vec{ v } ) = \frac { 1} { \left( 2\pi \sigma _ { v } ^ { 2} \right) ^ { 3/ 2} N _ { \text{esc} } } \exp \left( - \frac { \left( \vec{ v } + \vec{v}_\odot \right) ^ { 2} } { 2\sigma _ { v } ^ { 2} } \right)\quad \text{for } v \leq v_\mathrm{max}\,,
\end{align}
where the velocity dispersion of the halo is roughly $\sigma_v \approx 156 \,\mathrm{km} \,\mathrm{s}^{-1}$ \cite{Kerr:1986hz,Green:2011bv} and  the normalization constant is:
\begin{align}
N _ { \text{esc} } = \operatorname{erf} \left( \frac { v _ { \text{esc} } } { \sqrt { 2} \sigma _ { v } } \right) - \sqrt { \frac { 2} { \pi } } \frac { v _ { \text{esc} } } { \sigma _ { v } } \exp \left( - \frac { v _ {\mathrm{esc}} ^ { 2} } { 2\sigma _ { v } ^ { 2} } \right)\,.
\end{align}
As we discussed in the introduction, we aim to remain agnostic about the `true' form of the local velocity distribution and we will therefore explore the impact of deviations from the Maxwell-Boltzmann form of Eq.~\eqref{eq:f_MB}.

Direct detection experiments aim to detect dark matter-induced nuclear recoils inside the target material. The expected rate of recoil events on the nuclei $N_i$ is calculated from (see {\it e.g.} \cite{Cerdeno:2010jj})
\begin{align}
\frac{\text{d} R_i}{dE_R}=\frac{\xi_i\;\rho_\text{loc}}{\mdm\;m_{N_i}}\int_{v^{(\text{D})}\geq v_\text{min,i}^{(\text{D})}(\ER)} \text{d}^3v^{(\text{D})}\, v^{(\text{D})}\,f(\vec{v}^{(\text{D})}+\vecvobs(t))\,\frac{\text{d}\sigma_i}{dE_R} (v^\text{(D)},E_R)\,,
\end{align}
where $\vec v^{(\text{D})}$ is the DM velocity in the rest frame of the detector and $\vecvobs$ is the velocity of the observer with respect to the Sun (for which we use the parametrization given in \cite{McCabe:2013kea}). In addition, $\text{d}\sigma_i/\dER$ is the differential scattering cross section of a dark matter particle with a nucleus of mass $m_{N_i}$ and mass fraction $\xi_i$ inside the detector. A dark matter particle must have a velocity larger than $v_{\text{min},i}^{(\text{D})}(\ER) = \sqrt{m_{N_i}\ER/(2\mu_{N_i}^2)}$ in order to transfer energy $\ER$ onto the nucleus, where $\mu_{N_i}$ is the reduced mass of the dark matter-nucleus system. Finally, the total rate is calculated from integrating the differential scattering rate over all possible nuclear recoil energies and summing over all nuclei, appropriately weighting by the probability of detecting a recoil of the nucleus $N_i$ with energy $\ER$,  $\epsilon_i(\ER)$:
\begin{align}
R = \int_{0}^{\infty}\,\sum_i \epsilon_i(E_R)\,\frac{\text{d}R_i}{\text{d} E_R}\,\text{d} E_R\,.
\end{align}
The expected number of recoil events during the exposure of the experiment, ${\cal E}$, can be calculated from the recoil rate via $\mathcal{N}=R {\cal E}$.
 
Dark matter particles traversing the Sun can lose energy due to scatterings with nuclei in the Solar interior and become gravitationally bound to it. 
Captured dark matter particles will eventually sink to the core, generating an overdensity of dark matter particles where the rate of annihilations can be sufficiently large to produce an observable neutrino flux at Earth \cite{Silk1985}.
We assume that capture and annihilation of dark matter particles are in equilibrium, thus the annihilation rate $\Gamma_a$ (and therefore the neutrino flux) from the Sun is completely determined by the capture rate, $\Gamma_a = C/2$, where the capture rate is given by \cite{Gould:1987ir,Gould1992}
\begin{align}
C=\sum_i \int_0^{R_\odot} 4\pi\,r^2\,\text{d} r \, \eta_i(r)\,&\frac{\rho_\text{loc}}{\mdm}\,\int_{v \leq v_{\text{max},i}^{\text{(Sun)}}(r)} \text{d}^3 v \, \frac{ f (\vec{v})}{v}\,w^2(r) \int_{\mdm v^2 /2}^{2 \mu_{N_i}^2 w^2(r)/\mNi} \dER \, \frac{\text{d} \sigma_i}{\dER}(w(r), \ER) \,.
\label{eq:general_formula_capture_rate}
\end{align}
Here, $v$ is the DM velocity in the rest frame of the sun and $w(r)\equiv (v^2+v^2_{\rm esc})^{1/2}$.
Furthermore, $\eta_i(r)$ is the number density profile of element $i$ in the sun, for which we adopt the Solar model AGSS09 \cite{Serenelli:2009yc}.
We denote the escape velocity from the sun at distance $r$ from the center by $v_\text{esc}(r)$ and $R_\odot$ is the Solar radius.
Finally, the maximal velocity for which capture inside the sun is possible is given by $v_{\text{max},i}^{\text{(Sun)}}(r) = 2 \, v_\text{esc}(r) \sqrt{\mdm \mNi}/\left| \mdm - \mNi\right|$.

Direct dark matter searches are subject to major uncertainties due to our ignorance of the interactions of dark matter particles with nuclei, in addition to our ignorance of the local dark matter density and velocity distribution.
One common approach is to express the differential scattering cross section in terms of spin-independent (SI) and spin-dependent (SD) interactions \cite{Cerdeno:2010jj}:
\begin{align}
\frac{\text{d}\sigma_i}{\dER}(\hat v, E_R)=\frac{\mNi}{2\,\mu_{N_i}^2\,\hat{v}^2}\,(\sigma_\text{SI}\,F_{\text{SI},i}^2(\ER)+\sigma_\text{SD}\,F_{\text{SD},i}^2(\ER)) \,,
\label{eq:x-section}
\end{align}
where $\sigma_\text{SI}$ and $\sigma_\text{SD}$ are the SI and SD scattering cross sections at zero momentum transfer,
$\hat v$ denotes the dark matter velocity relative to the target nucleus, while $F_{\text{SI},i}(E_R)$ and $F_{\text{SD},i}(E_R)$ are form factors that depend on the structure of the nucleus. In this work, we adopt the standard SI Helm form factor \cite{Lewin1996}, and the SD form factors reported in \cite{Klos:2013rwa} and \cite{Catena:2015uha} relevant for direct detection experiments and neutrino telescopes,  respectively. In the following, we treat the dark matter mass $\mdm$ and the SI (respectively SD) scattering cross section at zero momentum transfer as free parameters. It is also common to assume a local dark matter density of $\rholoc=0.3~\GeVcm$ and a Maxwell-Boltzmann velocity distribution, as described at the start of this section. Under these assumptions, null search results from dark matter experiments can be translated into a limit on the SI or SD cross section at zero momentum transfer as a function of the dark matter mass.

\section{Optimizing the dark matter distribution}
\label{sec:optimization}

In this work, we investigate how the limits on dark matter parameters are affected if the true velocity distribution inside the Solar system departs from the Maxwell-Boltzmann form, Eq.~\eqref{eq:f_MB}. To this end, we require that at any point in velocity space the relative difference between the true velocity distribution and the Maxwell-Boltzmann distribution is at most a factor $\Delta$:
\begin{align}
\left|\frac{f(\vec v)-f_{\rm MB}(\vec v)}{f_{\rm MB}(\vec v)}\right|\leq \Delta\,.
\label{eq:def_Delta}
\end{align}
The phenomenological parameter $\Delta$ thus allows us to quantify the deviation of the velocity distribution from the SHM form. We note that in our analysis, $\Delta$ is independent of $\vec{v}$, such that the constraint in Eq.~(\ref{eq:def_Delta}) applies across the entire velocity distribution, for a single, constant value of $\Delta$. Note also that for $\Delta\geq 1$ the velocity distribution may vanish in parts of the velocity space (though it cannot become negative), whereas for $\Delta<1$ the velocity distribution is necessarily non-vanishing in the whole range.  We then calculate, for a given $\Delta$, the velocity distribution that maximizes/minimizes the signal rate in a dark matter search experiment. Correspondingly, we calculate the most conservative/most aggressive limits on the cross section as a function of the mass, compatible with the condition Eq.~(\ref{eq:def_Delta}).

More specifically, for a given dark matter mass, cross section and velocity distribution  we calculate  the $p$-value from the cumulative Poisson probability distribution of obtaining the observed number of events, or less, given an expected number $\NBG$ of background events and a number $\Nsig$ of expected signal events
\begin{align}
p_A(\Nsig) &= P( k \leq \Nobs | \NBG + \Nsig)\,.
\end{align}
Then, we optimize the $p$-value for a given experiment after sampling over velocity distributions, with the constraint Eq.~(\ref{eq:def_Delta}) as well as the normalization condition Eq.~(\ref{eq:VDNormalization}). Namely,
\begin{align}
\begin{split}
\text{Optimize:}~~~~~&\log p\,(\mdm,\sigma)\label{eq:Optimization_Problem}\\
\text{Subject to:}~~~~~
&\left|\frac{f(\vec v)-f_{\rm MB}(\vec v)}{f_{\rm MB}(\vec v)}\right|\leq \Delta \\
\text{and}~~~~~ & \int \text{d}^3 vf(\vec v)=1\,.
\end{split}
\end{align}
Finally, we determine,  for a given DM mass, the most conservative 90\% limit on the cross section from the condition
\begin{align}
\max_{f(\vec v)} \left\{\log p_\mathrm{tot} \right\}(\mdm, \sigma) \geq \log (0.1)\,,
\end{align}
while the most aggressive  90\% limit on the cross section is obtained from
\begin{align}
\min_{f(\vec v)} \left\{\log p_\mathrm{tot} \right\}(\mdm, \sigma) \geq \log (0.1)\,.
\end{align}

In order to perform the optimization, we discretize the velocity distribution as:
\begin{align}
f(\vec v)=\sum c_{\vec v_i} \delta(\vec v-\vec v_i)\,.
\end{align}
Then, the optimization problem over a continuous function can be recast as an optimization problem over a finite number of variables. In our numerical analysis, we use a total of 3000 streams, linearly spaced in the range $v \in [0, \,v_\mathrm{max}]$. The problem is then reduced to an optimization problem over the stream densities $c_{\vec v_i}$:
\begin{align}
\begin{split}
\text{Optimize:}~~~~~
&{\log p}\,(c_{\vec v_i},\mdm,\sigma)  \\
\text{Subject to:}~~~~~
&c_{\vec v_i}\leq (\Delta+1)f_{\rm MB}(\vec v_i)  \\
&c_{\vec v_i}\geq {\rm max}\Big\{0,(1-\Delta)\Big\}f_{\rm MB}(\vec v_i)  \\
&\sum_i c_{\vec v_i}=1\,. \label{eq:Optimization_Problem_Discrete}
\end{split}
\end{align}

This formalism -- decomposing the velocity distribution as a large number of streams and optimizing over the stream densities -- is completely general and can be applied to arbitrary likelihoods. However, with such a large number of poorly constrained parameters $c_{\vec v_i}$ this would typically be slow (and in some cases intractable) using, for example, Monte Carlo methods. If the value of $\log p$ in Eq.~\eqref{eq:Optimization_Problem_Discrete} were linear in the number of signal events (and therefore linear in the stream densities $c_{\vec v_i}$), the optimization problem can be solved quickly with the linear optimization techniques of IR17. As we will see in Sec.~\ref{sec:DD}, this approximation holds reasonably well in the case of XENON1T-2017 and PICO. More generally, the log-$p$-value (or log-likelihood) will exhibit a more complicated dependence on the $c_{\vec v_i}$. However, as we describe in Sec.~\ref{sec:NT} and Sec.~\ref{sec:DD+NT}, we can typically write down $\log p$ (for example) as quadratic in the $c_{\vec v_i}$ parameters, to a good approximation. This allows us to make use of quadratic programming techniques (in particular CVXOPT's optimizer for quadratic programs \cite{CVXOPT,Boyd2004}) to efficiently solve the optimization problem and determine the most aggressive and most conservative limits over all possible velocity distributions.

The normalization of the DM signal also depends on the local DM density $\rho_\mathrm{loc}$. We therefore report our results in terms of $\sigma \cdot \rho_{\operatorname { loc } _ { 0.3 }}\equiv \sigma \cdot (\rho_\mathrm{loc}/(0.3 \,\mathrm { GeV } / \mathrm { cm } ^ { 3 }))$, factorizing out the dependence of the limits on $\rho_\mathrm{loc}$.  A larger (or smaller) value of the local density would therefore simply rescale our limits and contours upwards (downwards). In addition, we include intrinsic uncertainties in the SHM distribution itself, through $v_\odot$  and $v_\mathrm{esc}$. In concrete models of the Milky Way $v_\odot$, $v_\mathrm{esc}$ and $\rho_\mathrm{loc}$ will be correlated \cite{Benito:2016kyp}. Remaining agnostic of specific Milky Way mass models, however, we treat these parameters as independent. In all cases, we vary $v_\odot$ in the range 220--240 ${\rm km}/{\rm s}$, $v_{\rm esc}$ in the range 499--608 ${\rm km}/{\rm s}$, and the angle between the detector velocity and the Sun's velocity between 0 and $2\pi$. 
We then determine the most aggressive and conservative limits within those ranges, as a function of $\Delta$.  Using the optimization techniques described above, this allows us to quickly and efficiently explore the full spectrum of possible astrophysical uncertainties.

We discuss in the next section some applications of our procedure to bracket the impact of our ignorance of the velocity distribution on {\it i)} limits on the cross section from direct detection experiments,  {\it ii)} limits from neutrino telescopes,   {\it iii)}  combined limits from direct detection and from neutrino telescopes,  {\it iv)} the reconstruction of parameters in view of a future putative signal. 

\section{Applications}
\label{sec:Applications}

\subsection{Impact on limits from direct detection experiments}
\label{sec:DD}

We first derive conservative and aggressive upper limits on the dark matter interaction cross section with nuclei following the method described in the previous section. To this end, we will use the null search results from XENON1T (2017 exposure) \cite{Aprile:2017iyp} and from PICO-60 \cite{Amole:2017dex} for the SI and the SD interaction, respectively\footnote{During the preparation of this manuscript, the XENON1T collaboration released results from a 1 tonne-year exposure \cite{Aprile:2018dbl}, strengthening constraints on spin-independent cross sections by a factor of between 2 and 5 (depending on the DM mass). In this work, we do not consider the more recent exposure but our methods can be applied straightforwardly to that setup. Unless otherwise stated, we use `XENON1T' to refer to the 2017 exposure.}. In both cases, the observed number of events in the signal region is zero, therefore the Poisson probability for obtaining the observed number of events, or less, is given by
\begin{equation}
\log(p)=-(N_\text{BG}+N_\text{sig}),
\end{equation}
where $N_\text{sig}$ is the expected number of signal events and $N_\text{BG}$ is the expected number of background events (where we neglect background uncertainties). For the case of XENON1T, $N_\text{BG}=0.36$
while for PICO-60, $N_\text{BG}=0.331$. To calculate the number of signal events we use the efficiency given in  \cite{Workgroup:2017lvb,DDcalc} for the XENON1T experiment and in \cite{Amole:2015lsj} for the PICO experiment. 

\begin{figure}[t!]
	\begin{center}
		\hspace{-0.75cm}
		\includegraphics[width=0.5\textwidth]{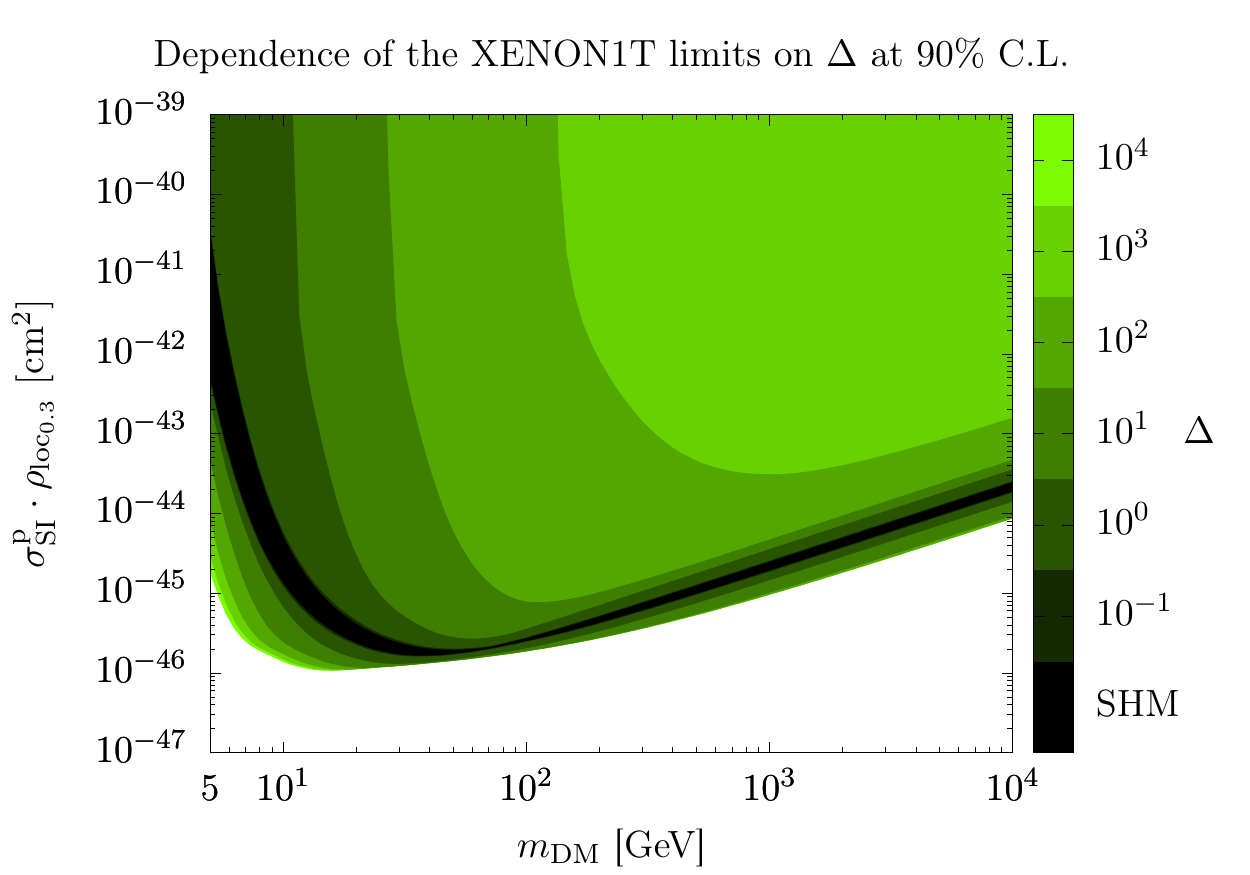}
		\includegraphics[width=0.5\textwidth]{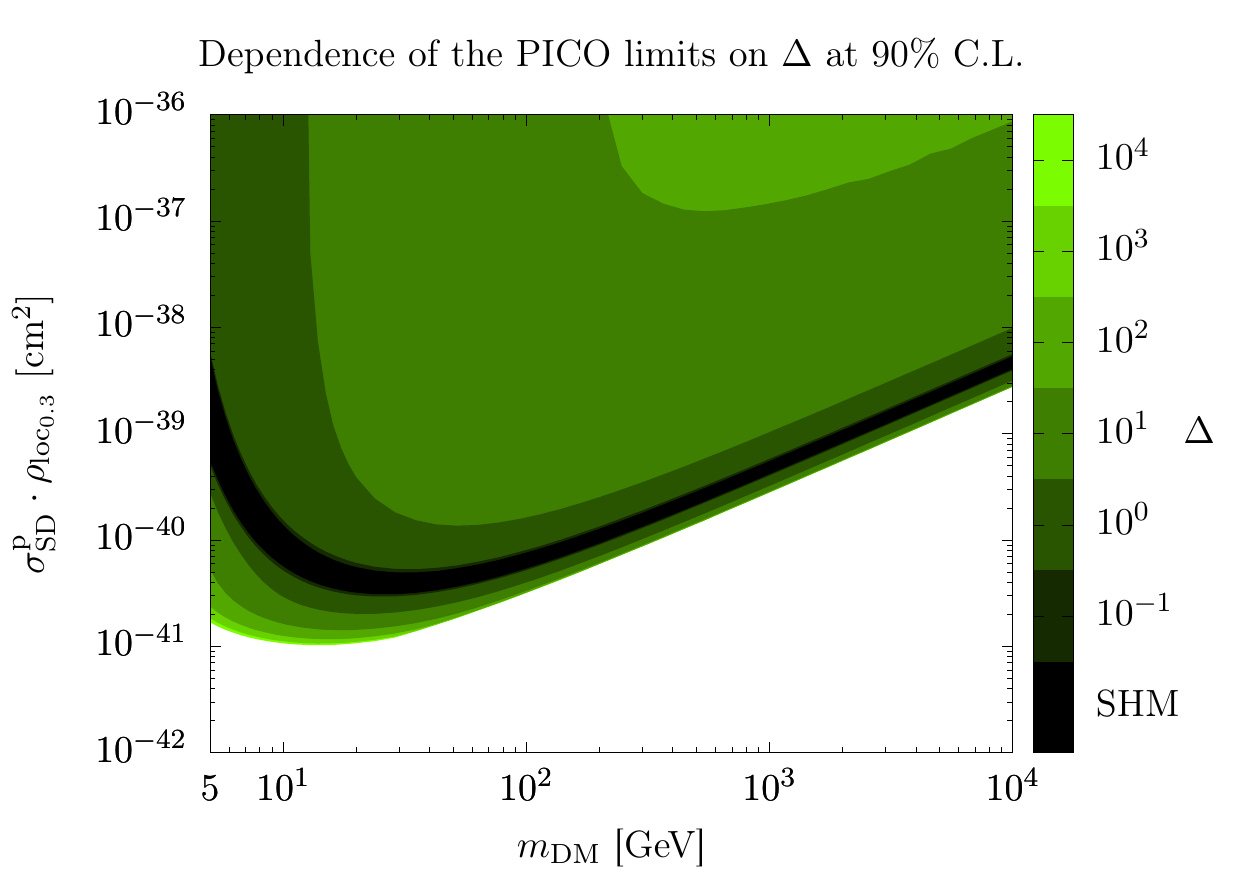}
	\end{center}
	\caption{Conservative and aggressive limits on the SI (left panel) and SD (right panel) interaction cross section from the null search results from XENON1T and PICO-60, respectively. Deviations from the Maxwell-Boltzmann distribution are parametrized through $\Delta$, defined in Eq.~\eqref{eq:def_Delta}. The limits include uncertainties in the determination of the parameters of the Maxwell-Boltzmann distribution (see Sec.~\ref{sec:optimization} for details). }\label{fig:DDonly}
\end{figure}

We show in Fig.~\ref{fig:DDonly} our conservative and aggressive limits on the SI (left panel) and SD (right panel) scattering cross sections for various choices of $\Delta$ between $0$ (corresponding to the SHM) and $10^4$. In the case of the SHM, we also allow $v_\odot$ and $v_\mathrm{esc}$ to vary (as described in the previous section) and select the most aggressive and conservative limits in each case. Allowing full freedom in the velocity distribution, the most aggressive limit on the SI (SD) cross section is, for large masses, at most a factor $\sim 2$ ($\sim 1.5$) stronger than the limit obtained from the SHM, while for small masses the limits can be significantly strengthened. The velocity distribution giving the largest scattering rate corresponds to 
\begin{align}
f_{\rm R_{max}}(v)=& 
f_{\rm MB}(v)\Big[(1+\Delta)\Theta(v-v_1)\Theta(v_2-v) \nonumber\\&\qquad+ (1-\Delta)\Theta(1-\Delta)  \Big(1-\Theta(v-v_2)\Theta(v_1-v)\Big)\Big]\,,
\label{eq:f_Rmax-DD}
\end{align}
where $v_1$ and $v_2$ are labeled such that $v_1<v_2$ and are calculable for a given dark matter mass and $\Delta$. Here, $\Theta(x)$ is the Heaviside function, defined as $\Theta(x)=1$ for $x\geq 0$ and $\Theta(x)=0$ otherwise. The high and low energy tails of the distribution are depleted of dark matter particles, which instead populate the velocity range $v_1<v<v_2$ where the scattering rate maximizes, in such a way that the conditions Eq.~(\ref{eq:VDNormalization}) and Eq.~(\ref{eq:def_Delta}) are fulfilled. Note that for $\Delta\gg 1$ the normalization condition translates into
\begin{align}
\int_{v_1}^{v_2} \text{d}v \,f_{\rm MB}(v) \simeq \frac{1}{\Delta}\,,
\end{align}
therefore, as $\Delta$ increases, the support of the function $f_{\rm R_{max}}(v)$ decreases, and tends in the limit $\Delta\rightarrow \infty$ to a $\delta$-function at the velocity which maximizes the recoil rate. This is to be expected, since for very large $\Delta$ the velocity distribution is in practice constrained only by the normalization condition. Hence, following the general discussion in IR17, the optimized velocity distribution must consist of just one dark matter stream. For DM masses larger than $\sim 100$ GeV, the speed of this stream is comparable to the mean velocity in the MB distribution, thus leading to only a modest strengthening of the limits. However, as the DM mass decreases, the speed of the stream shifts to larger and larger values, so the DM momentum increases and makes nuclear recoils above the detectable threshold possible. In this regime, the optimized velocity distribution is significantly different to the MB distribution (which is exponentially suppressed at high $v$) and accordingly the limits on the cross section can be notably strengthened. Clearly, since the speed of the stream is bounded from above by $v_{\rm max}$, the experiment becomes insensitive to DM scatterings for sufficiently small masses.

The most conservative limits, on the other hand, can be much weaker than those derived from the SHM. The velocity distribution giving the smallest scattering rate corresponds to 
\begin{align}
f_{\rm R_{min}}(v)=
f_{\rm MB}(v)\Big[&(1+\Delta)\Big(\Theta(v'_1-v)+\Theta(v-v'_2)\Big) + \nonumber \\ &(1-\Delta)\Theta(1-\Delta)\Big(\Theta(v-v'_1)+\Theta(v'_2-v)-1\Big)\Big]\,,
\label{eq:f_Rmin-DD}
\end{align}
where again $v'_1$ and $v'_2$ are calculable for a given dark matter mass and $\Delta$. In this case, the intermediate velocity range $v'_1<v<v'_2$ is depleted of dark matter particles, which instead populate the
high and low energy tails of the distribution, in such a way that the conditions Eq.~(\ref{eq:VDNormalization}) and Eq.~(\ref{eq:def_Delta})  are fulfilled. The qualitative features of $f_{\rm R_{min}}(v)$ are opposite to those of $f_{\rm R_{max}}(v)$. For $\Delta\gg 1$, the normalization condition implies
\begin{align}
\int_0^{v'_1} \text{d}v f_{\rm MB}+\int_{v'_2}^{v_{\rm max}}\text{d}v f_{\rm MB}\simeq\frac{1}{\Delta}\,.
\end{align}
In the limit $\Delta\rightarrow\infty$, one finds $v'_2>v_{\rm max}$ and $v'_1\rightarrow 0$, such that the velocity distribution reads $f_{\rm R_{min}}(v)\rightarrow \delta(v)$. In this limit, therefore, the scattering rate is zero for all dark matter masses, and correspondingly the scattering cross-section is unconstrained. As $\Delta$ decreases, the velocity distribution $f_{\rm R_{min}}$ takes non-vanishing values for velocities inducing observable recoils and accordingly the minimum value of the scattering rate is non-vanishing. As a result, the most conservative limits on the scattering cross section vary between infinity, for $\Delta\rightarrow\infty$, and the SHM value, for $\Delta\rightarrow 0$.

\subsection{Impact on limits from neutrino telescopes}
\label{sec:NT}

To determine the limits on the scattering cross section from neutrino telescopes we use respectively the 3 year data sample of the neutrino flux from  IceCube (IC) or from  DeepCore (DC)\cite{Aartsen:2016zhm}, for DM masses above or below 100 GeV. The full IceCube analysis uses an event-by-event likelihood, but the collaboration also present binned data as a function of the direction of the event, measured as the angle $\psi$ from the direction of the Sun. For the low-energy DC sample we include all 7 angular bins in the analysis, while for the IC sample, we use only the 3 angular bins pointing closest to the Sun's position, due to the improved angular resolution of the signal in this case. 

We construct a $p$-value based on a single-bin counting experiment in each of the two samples (DC or IC). To account for uncertainties in the background expectation, we sum the background uncertainties in quadrature (for all the bins we include) and then allow for a downward fluctuation of $1\sigma$ in the total background expectation. This allows us to set conservative upper limits on the DM-nucleon cross section without performing a full likelihood analysis. The number of events in each sample then are as follows:
\begin{align}
\begin{split}
\label{eq:ICdata}
\text{DeepCore (7 bins):}& \qquad \Nobs = 427 \qquad \NBG (-1\sigma) = 414\,,\\
\text{IceCube (3 bins):}& \qquad \Nobs = 926 \qquad \NBG (-1\sigma)  = 931\,.
\end{split}
\end{align}

The $p$-value will be obtained as before, from the cumulative Poisson probability distribution:
\begin{align}
p_A(\Nsig) &= P( k \leq \Nobs | \NBG + \Nsig)\,,
\end{align}
where $A = \mathrm{IC},\,\mathrm{DC}$ depending on the DM mass. 
Setting $p = 0.1$ for a 90\% upper limit gives $\Nsig^{90\%} = 35.0$ for IceCube and $\Nsig^{90\%} = 39.9$ for DeepCore, in rough agreement with the limits given in Table~4 of Ref.~\cite{Aartsen:2016zhm}. However, we note that using a simple counts-based approach to setting limits (rather than a full likelihood) tends to strengthen the limits slightly at low DM mass and weaken the limits at high DM mass \cite{Aartsen:2016exj}. We approximate $\log(p_A)$ as a second order polynomial:
\begin{align}
\begin{split}
\log(p_\mathrm{DC}) &= -0.319 - 0.0155 \,\Nsig - 0.0008 \,\Nsig^2\,,\\
\log(p_\mathrm{IC}) &= -0.823 - 0.0276 \,\Nsig - 0.0004 \,\Nsig^2\,.
\end{split}
\end{align}
In this case, the fitting functions are accurate to within 10\% for $\Nsig \in [0, 10]$, improving to 1\% accuracy for $\Nsig \in [10, 100]$. Finally, we relate the number of signal events to the capture rate $\Gamma_C$ (or equivalently the annihilation rate $\Gamma_\mathrm{ann}$) following closely Ref.~\cite{Aartsen:2016zhm}.

\begin{figure}[t!]
	\begin{center}
		\hspace{-0.75cm}
		\includegraphics[width=0.49\textwidth]{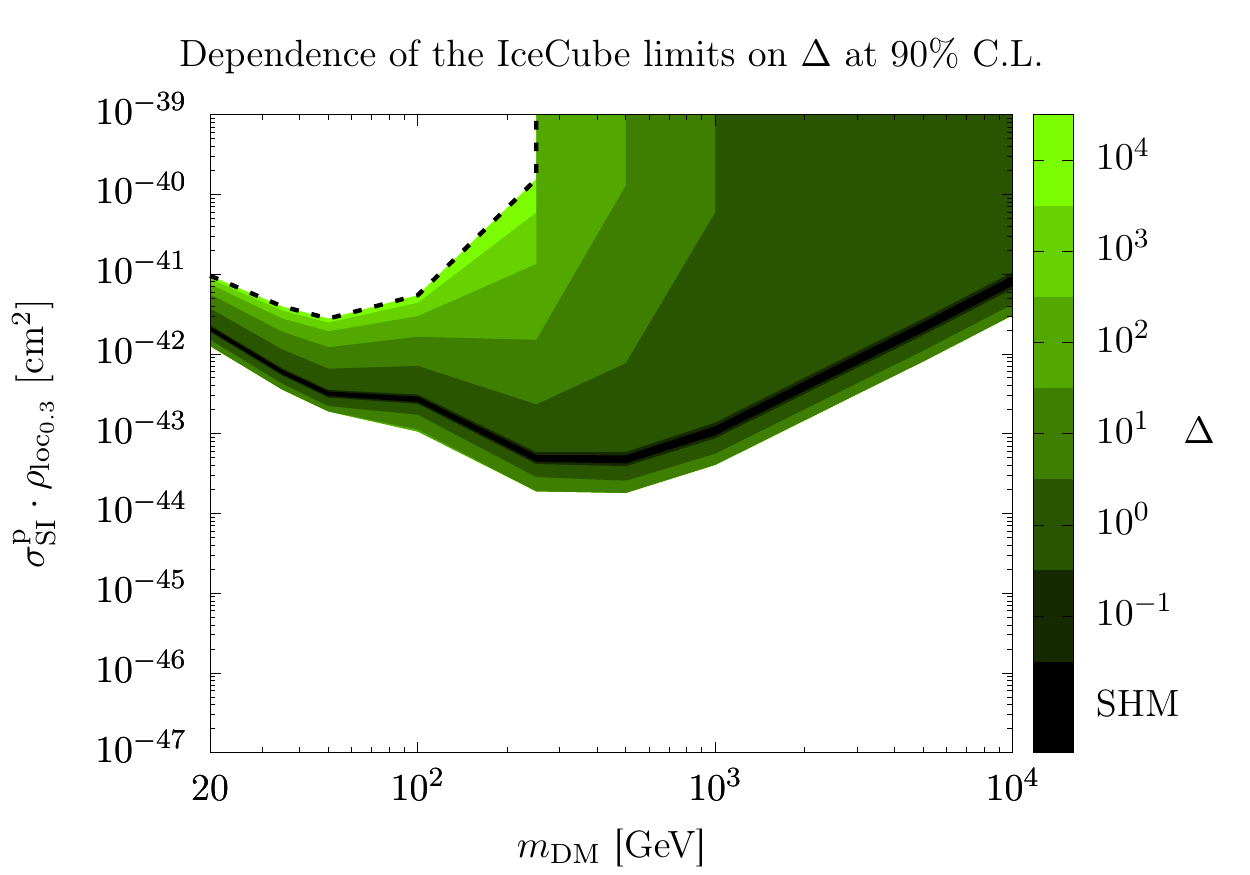}
		\includegraphics[width=0.49\textwidth]{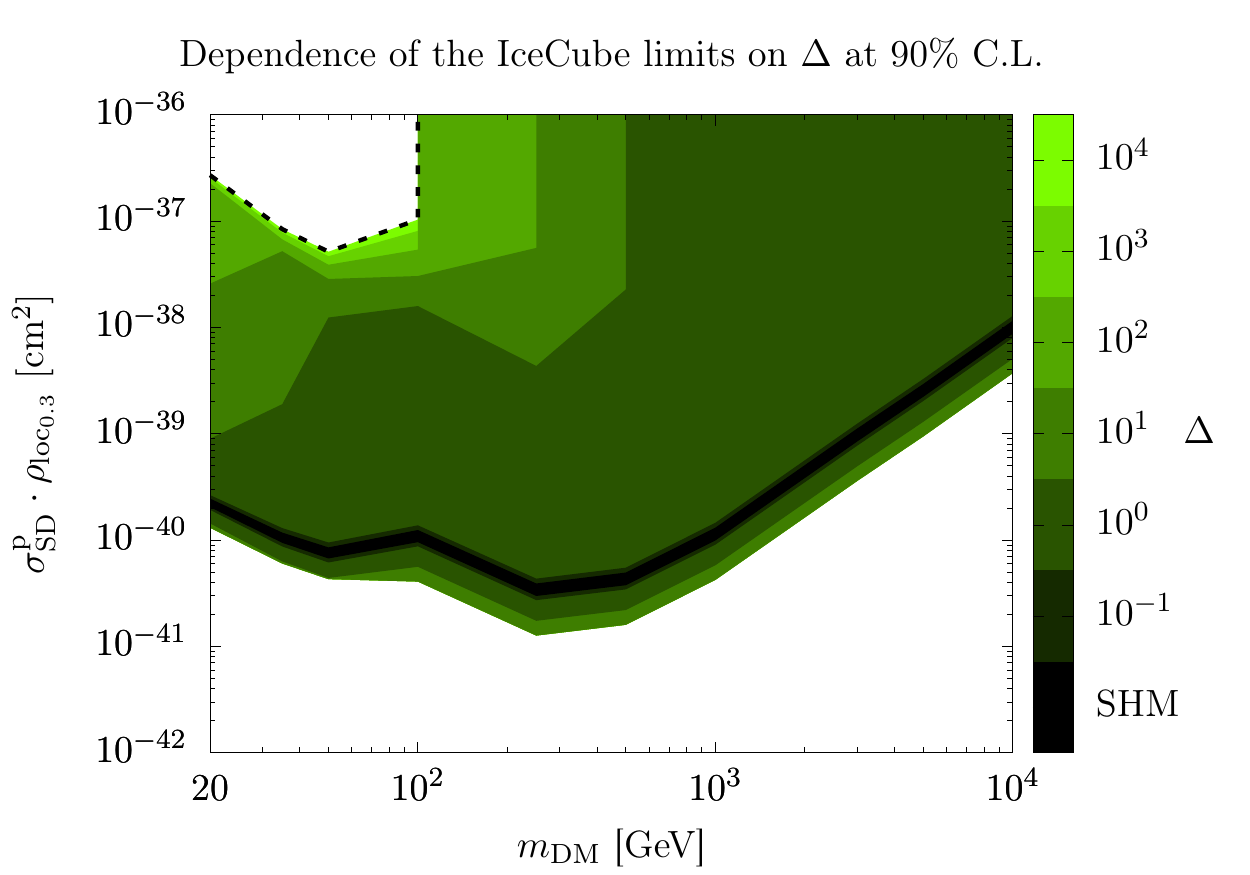}
	\end{center}
	\caption{Same as Fig.~\ref{fig:DDonly} but for the null results from IceCube and DeepCore. }\label{fig:IceCubeOnly}
\end{figure}

In Fig.~\ref{fig:IceCubeOnly} we show the 90\% C.L. aggressive and conservative limits on the SI (left panel) and the SD (right panel) interaction cross section from the null search results from IceCube and DeepCore for $\Delta$ between 0 and $10^4$, using the approach in Eq.~(\ref{eq:Optimization_Problem}), and assuming dark matter annihilations into $W^+W^-$ (or $\tau^+\tau^-$ for DM masses below the $W$ threshold of 80 GeV). We also assume equilibrium between dark matter capture and annihilation\footnote{We have explicitly checked that this is a good approximation for our optimized velocity distributions for the values of the cross-section that saturate the limits, if the annihilation cross section is equal to the thermal value.}.  We find that the most aggressive limits on the cross section are at most a factor 2.5 more stringent than the SHM limit, whereas the most conservative limits can be significantly weaker than those from the SHM, especially for large DM masses.

The velocity distribution giving the largest capture rate corresponds to 
\begin{align}
f_{\rm C_{max}}(v)&= f_{\rm MB}(v)\Big[(1+\Delta)\Theta(v_1-v) + (1-\Delta)\Theta(1-\Delta)\Theta(v-v_1)\Big]
\label{eq:f_Cmax-DD}\,,
\end{align}
with $v_1$  calculable for a given dark matter mass and $\Delta$. This form is a consequence of the fact that the capture rate decreases monotonically with the dark matter velocity, so that the  velocity distribution providing the largest rate corresponds to that where the low velocity part of the distribution is as populated as possible, subject to the  constraints in  and Eq.~(\ref{eq:VDNormalization}) and Eq.~(\ref{eq:def_Delta}). Moreover, the velocity $v_1$ is now easily calculable from the normalization condition:
\begin{align}
\int_0^{v_1} f_{\rm MB}(v)\,\text{d}v=\frac{1-(1-\Delta)\Theta(1-\Delta)}{1+\Delta -(1-\Delta)\Theta(1-\Delta)}\,.
\end{align}
Clearly, as $\Delta$ increases, the support of the function $f_{\rm C_{max}}(v)$ is restricted to a smaller and smaller interval left-bounded by $v=0$, and tends to $f_{\rm C_{max}}(v)=\delta(v)$ as $\Delta\rightarrow \infty$. Again, this is a consequence of the fact that in this limit the optimization problem is only subject to the normalization constraint, and therefore the optimized velocity distribution is composed of a single delta-function~\cite{Ibarra:2017mzt}.

The velocity distribution giving the smallest capture rate corresponds to 
\begin{align}
f_{\rm C_{min}}(v)=& f_{\rm MB}(v)\Big[(1+\Delta)\Theta(v-v'_1) + (1-\Delta)\Theta(1-\Delta)\Theta(v'_1-v)\Big]\,,
\label{eq:f_Cmin-DD}
\end{align}
which populates the high velocity part of the distribution at the expense of the low velocity part. Again, the qualitative features of $f_{\rm C_{min}}(v)$ are opposite to those of $f_{\rm C_{max}}(v)$. The velocity $v'_1$ is calculable from the normalization condition (as in Eq.~\eqref{eq:f_Cmax-DD}):
\begin{align}
\int_{v'_1}^{v_{\rm max}} f_{\rm MB}(v)\,\text{d}v=\frac{1-(1-\Delta)\Theta(1-\Delta)}{1+\Delta -(1-\Delta)\Theta(1-\Delta)}\,.
\end{align}
In this case, as $\Delta$ increases, the support of the function $f_{\rm C_{min}}(v)$ is restricted to a smaller and smaller interval right-bounded by $v=v_{\rm max}$ and tends to $f_{\rm C_{min}}(v)=\delta(v-v_{\rm max})$ as $\Delta\rightarrow \infty$. In this regime, the capture rate is non-zero for small DM masses, meaning that the most conservative limit is finite, being bounded from above by the dashed line in Fig.~\ref{fig:IceCubeOnly}. This limit is weaker than the SHM limit by a factor of $\sim10 \, (1000)$ for SI (SD) interactions at low mass. For sufficiently large DM mass (250 GeV for SI and 100 GeV for SD) capture becomes no longer possible for a stream, as the scattering kinematics mean that the DM velocity after scattering is necessarily larger than the local Solar escape velocity. In this range of masses, the scattering cross section is therefore unconstrained by neutrino telescopes (for $\Delta \rightarrow \infty$).

\subsection{Impact on combined limits from direct detection experiments and neutrino telescopes}
\label{sec:DD+NT}

Various works have highlighted the complementarity of direct detection experiments in constraining halo-independently the dark matter properties, as these two search strategies probe different parts of the velocity space \cite{Kavanagh:2014rya,Ibarra:2017mzt,Blennow:2015oea,Ferrer:2015bta,Ferrer:2013cla}. This complementarity is apparent from the limit $\Delta\rightarrow \infty$ discussed in subsections \ref{sec:DD} and \ref{sec:NT}. The most aggressive  limits for direct detection experiments correspond to $f_{\rm R_{max}}(v)=\delta(v-v_1)$, with $v_1$ dependent on the DM mass, whereas  for neutrino telescopes to $f_{\rm  C_{max}}(v)=\delta(v)$. Conversely, the most conservative limits correspond to $f_{\rm R_{min}}(v)=\delta(v)$ and $f_{\rm  C_{min}}(v)=\delta(v-v_{\rm max})$ for direct detection experiments and for neutrino telescopes, respectively. Given the different qualitative features of the optimized velocity distributions in direct detection experiments and in neutrino telescopes, one expects a non-trivial complementarity between these two detection approaches also for the determination of the most aggressive and most conservative limits on the scattering cross section.

These limits can be calculated following Eq.~(\ref{eq:Optimization_Problem_Discrete}), with a total log-$p$-value given by the sum of those from the different experiments:
\begin{align}
\log p_\mathrm{tot} &= \sum_{k \in \mathrm{expts}} \log p_k\,,
\end{align} 
Here, $\log p_k$ is either linear in the number of direct detection events (for the case of PICO and XENON1T) or quadratic (for IceCube). Therefore, the total log-$p$-value is quadratic and one can implement the optimization method along the lines described in Sec.~\ref{sec:optimization}.

\begin{figure}[t!]
	\begin{center}
		\hspace{-0.75cm}
		\includegraphics[width=0.49\textwidth]{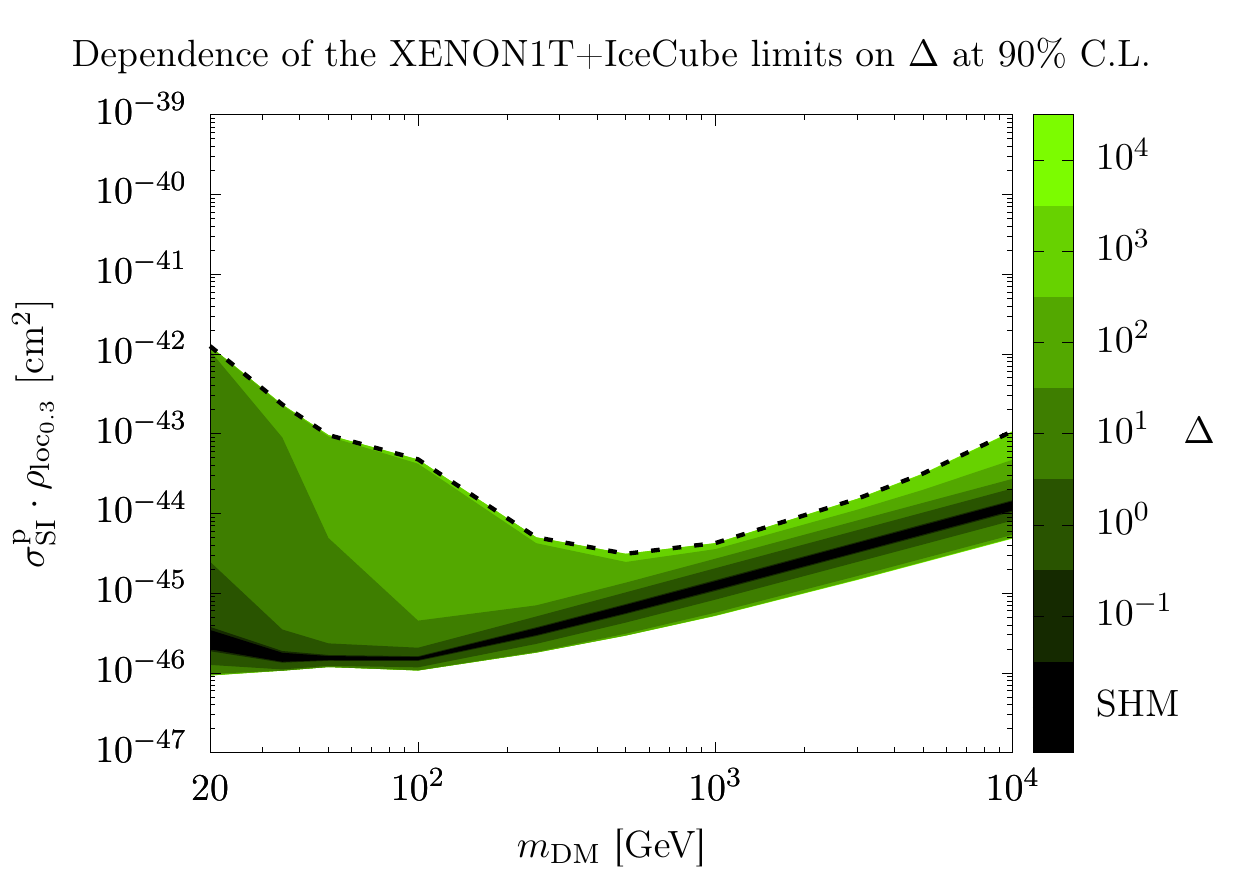}
		\includegraphics[width=0.49\textwidth]{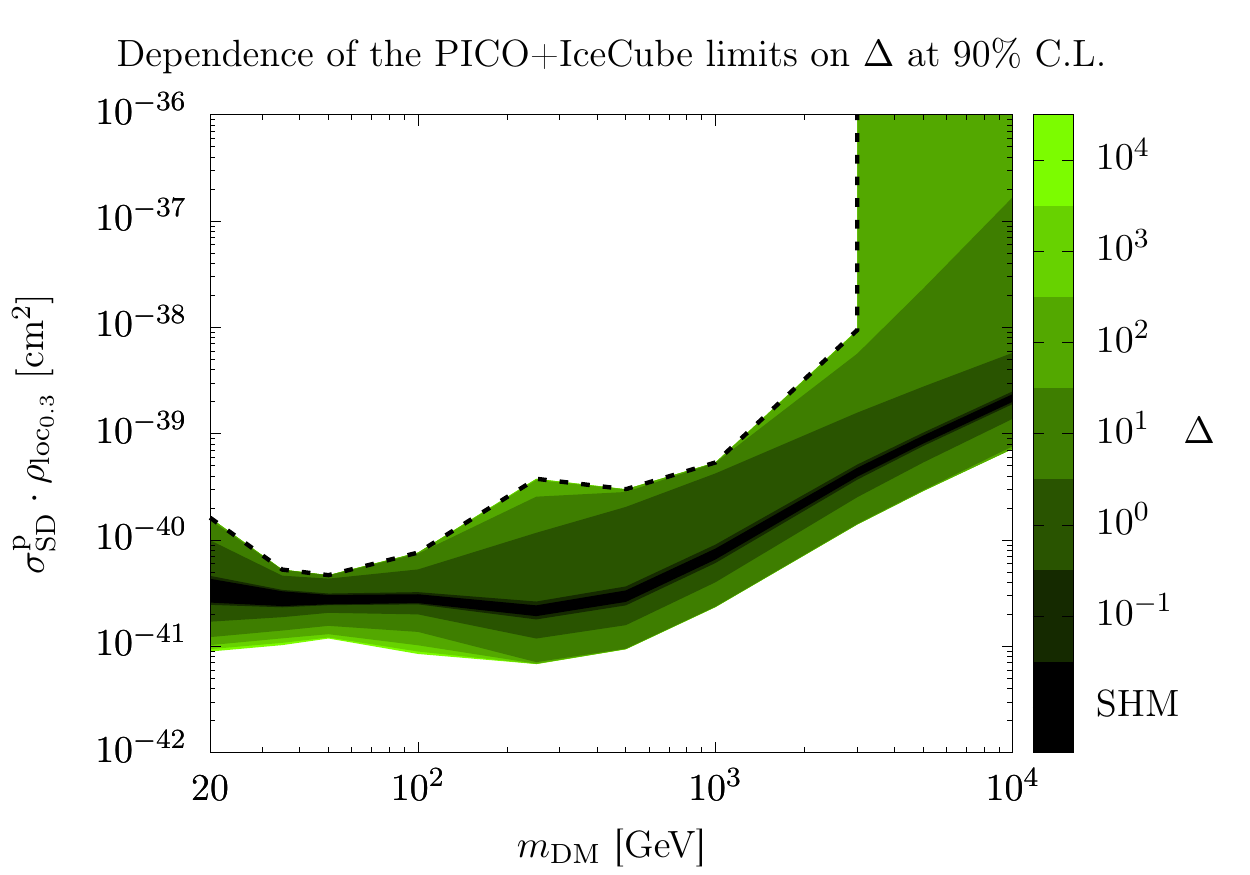}
	\end{center}
	\caption{Same as Fig.~\ref{fig:DDonly} but combining the null results from XENON1T and IceCube (left panel) and from PICO-60 and IceCube (right panel).}\label{fig:DDNT}
\end{figure}

We show in Fig.~\ref{fig:DDNT} the 90\% C.L. aggressive and conservative limits on the SI (left panel) and the SD (right panel) interaction cross section from combining, respectively, the null results of XENON1T and IC+DC, or PICO-60 and IC+DC, for different values of $\Delta$. The halo-independent aggressive limits are again not too different, at most a factor 2 (3 for SD) stronger, than those from the SHM, and correspond in the limit  $\Delta\rightarrow\infty$ to a superposition of at most two streams with velocities which can be calculated for a given dark matter mass, as discussed in IR17. In addition, the halo-independent conservative limits are, remarkably, not too different from those obtained for the SHM. Over the whole mass range analyzed for the SD interaction, and when $m_\mathrm{DM} \gtrsim 1\,\mathrm{TeV}$ for the SI interaction, the conservative halo-independent upper limits are a factor of 10 weaker than the SHM limits when $\Delta=10^4$, and a factor of 2 weaker when $\Delta=10$. Even allowing full freedom, the velocity distribution cannot be distorted to evade bounds from both direct detection and neutrino telescopes, probing as they do different velocity ranges. We note however that when  $m_{\rm DM}\lesssim 1$ TeV for the SI interaction, the conservative limits can be several orders of magnitude weaker than the SHM. This is because the IceCube bound is substantially weaker than the XENON1T bound at low mass; the XENON1T limits can be lifted by a large amount (by pushing DM particles to lower velocities) before the IceCube bound becomes relevant.

This analysis extends the one presented in  IR17  corresponding to the streams-only case in two respects: first, rather than combining the 90\% C.L. limits from each experiment to provide a combined upper limit, here we optimize the total $p$-value thus allowing us to rigorously assign a statistical significance to the combined upper limit. And second, we present a procedure to interpolate, in terms of the parameter $\Delta$, between the limits obtained in the SHM (corresponding to $\Delta=0$) and the most aggressive/conservative limits possible (corresponding to $\Delta\rightarrow\infty)$. With this technique we can bracket the limits on DM-nucleon interactions as a function of the SHM deviation $\Delta$. For each $\Delta$, the resulting aggressive and conservative limits are stronger than those provided by each experiment separately.

\subsection{Impact on parameter reconstruction}

In order to examine the prospects for reconstructing the DM properties from a future detection, we construct an approximate likelihood for a future version of the XENON1T experiment. We consider recoils in the range $E_R \in [3, 70]\;\mathrm{keV}$, using the nuclear recoil efficiencies from Refs.~\cite{Workgroup:2017lvb,DDcalc}. We consider a total exposure of $7.6 \times 10^5 \;\mathrm{kg}\,\mathrm{days}$, corresponding to a 1042 kg fiducial mass and a two year exposure. We divide the energy range into $N_\mathrm{bins}=4$ bins, chosen such that each bin is expected to contain at least 5 events, for a given DM mass and cross section. Neglecting backgrounds and systematic uncertainties, we estimate the uncertainty on the signal rate in each bin as $\sigma_i = (N_\mathrm{obs}^{(i)})^{1/2}$ where  $N_\mathrm{obs}^{(i)}$ is the observed number of signal events in the $i$th bin. Of course, in the event of a detection, we would be able to make use of $\sigma_i$ as estimated by the experimental collaboration, taking into account the appropriate systematic uncertainties. From this, we calculate the approximate Gaussian likelihood\footnote{Note that this definition of the likelihood differs slightly from the standard $\chi^2$ test-statistic for Poisson distributed data, which typically has $\sigma_i = (N_\mathrm{sig}^{(i)})^{1/2}$ instead of $\sigma_i = (N_\mathrm{obs}^{(i)})^{1/2}$, though in the signal dominated regime the two should give very similar results. The definition given in Eq.~\eqref{eq:likelihood} has the advantage that it is (at most) quadratic in the number of signal events and therefore the DM velocity distribution. We have explicitly checked that $X^2 = -2 \log\mathcal{L}$ approximately follows a $\chi^2$-distribution though typically with a slightly heavier tail. In addition, we have checked that confidence regions constructed using this test statistic should have roughly the correct coverage (within a few percent).},
\begin{equation}
\label{eq:likelihood}
\log\mathcal{L}_\mathrm{Xe} = -\frac{1}{2}\sum_i^{N_\mathrm{bins}} \frac{\left(N_\mathrm{sig}^{(i)} - N_\mathrm{obs}^{(i)}\right)^2}{\sigma_i^2}\,.
\end{equation}
Here, $N_\mathrm{sig}^{(i)}$ is the number of signal events expected in the $i$th bin, for a given model, specified by $\left(\mdm, \sigma_\mathrm{SI}^p, f(\vec{v})\right)$. 

We generate a mock data set consisting of so-called `Asimov' data \cite{Cowan:2010js}, setting $N_\mathrm{obs}^{(i)} = N_\mathrm{sig}^{(i)}$ for a given benchmark point in parameter space. As a benchmark, we generate data assuming $m_\mathrm{DM} = 50\,\mathrm{GeV}$ with a spin-independent cross section $\sigma^p_\mathrm{SI} = 1.5 \times 10^{-46}\,\mathrm{cm}^2$ and an underlying SHM distribution. We expect $\sim 34$ signal events at the benchmark point. We choose this point as an illustrative example, but the approach could easily be extended to other regions of parameter space using (for example) the benchmark-free forecasting techniques of Ref.~\cite{Edwards:2018lsl}.

The likelihood in Eq.~\eqref{eq:likelihood}  is quadratic in the number of signal events and therefore (at most) quadratic in the DM velocity distribution. We can therefore use the techniques of quadratic optimization to maximize the likelihood over all possible velocity distributions (or only a subset if desired). Confidence regions for the DM mass and cross section are then constructed in the standard way, profiling over velocity distributions and including all values of $\mdm$ and $\sigma_\mathrm{SI}^p$ which satisfy \cite{Wilks:1938dza,Cowan:2010js}:
\begin{equation}
\max_{f(\vec{v})} \left\{\log\mathcal{L}(\mdm, \sigma_\mathrm{SI}^p) \right\}>  \log\mathcal{L}_\mathrm{max} -  \frac{1}{2}\chi^2_{\gamma\%}\,.
\end{equation}
Here, $\log\mathcal{L}_\mathrm{max}$ is the maximum log-likelihood, maximized over all masses, cross sections and velocity distributions. The relevant critical $\chi^2$ value for 2 degrees of freedom, written as $\chi^2_{\gamma\%}$, depends on the confidence level of the contours. Here, we take $\chi^2_{90\%} \approx 4.605$.

\begin{figure}[t]
	\begin{center}
		\hspace{-0.75cm}
		\includegraphics[width=0.49\textwidth]{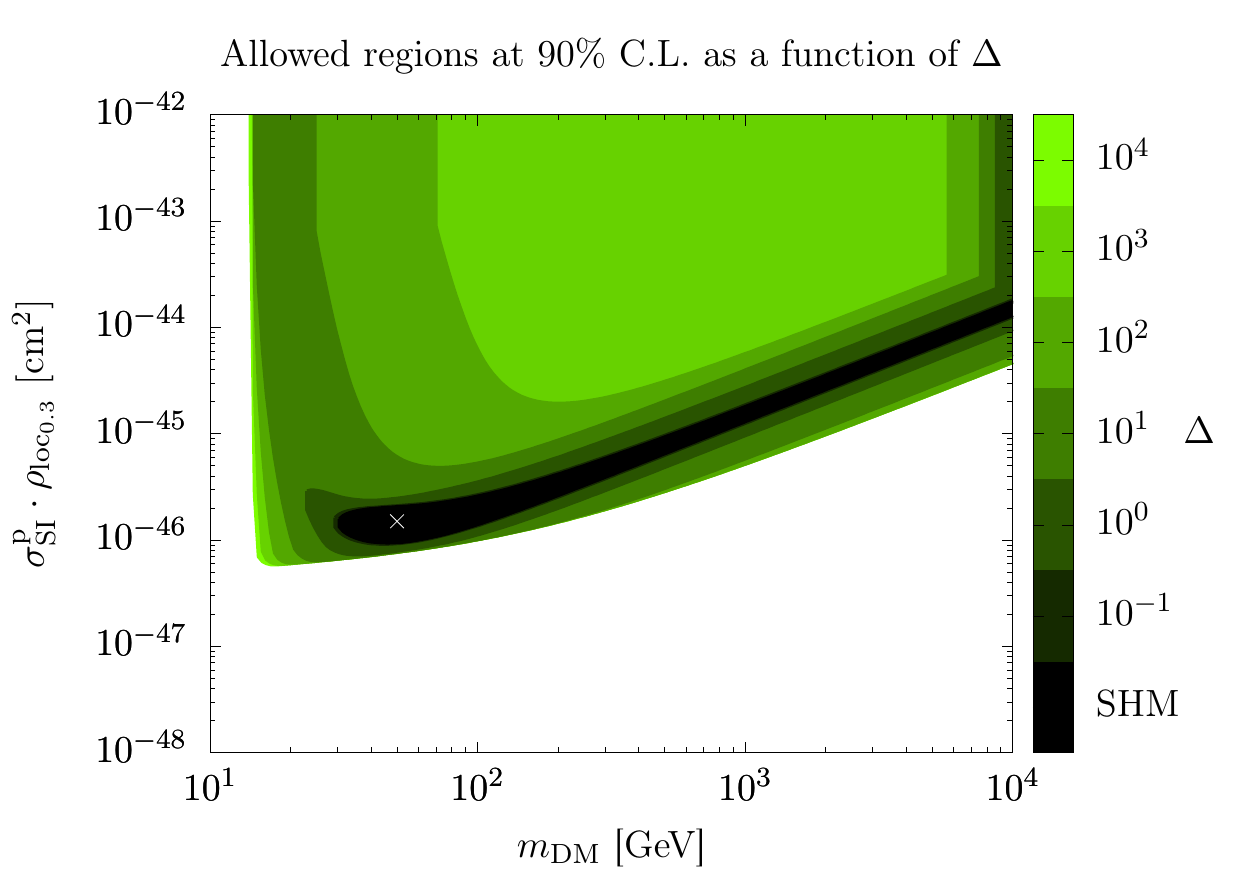}
		\includegraphics[width=0.49\textwidth]{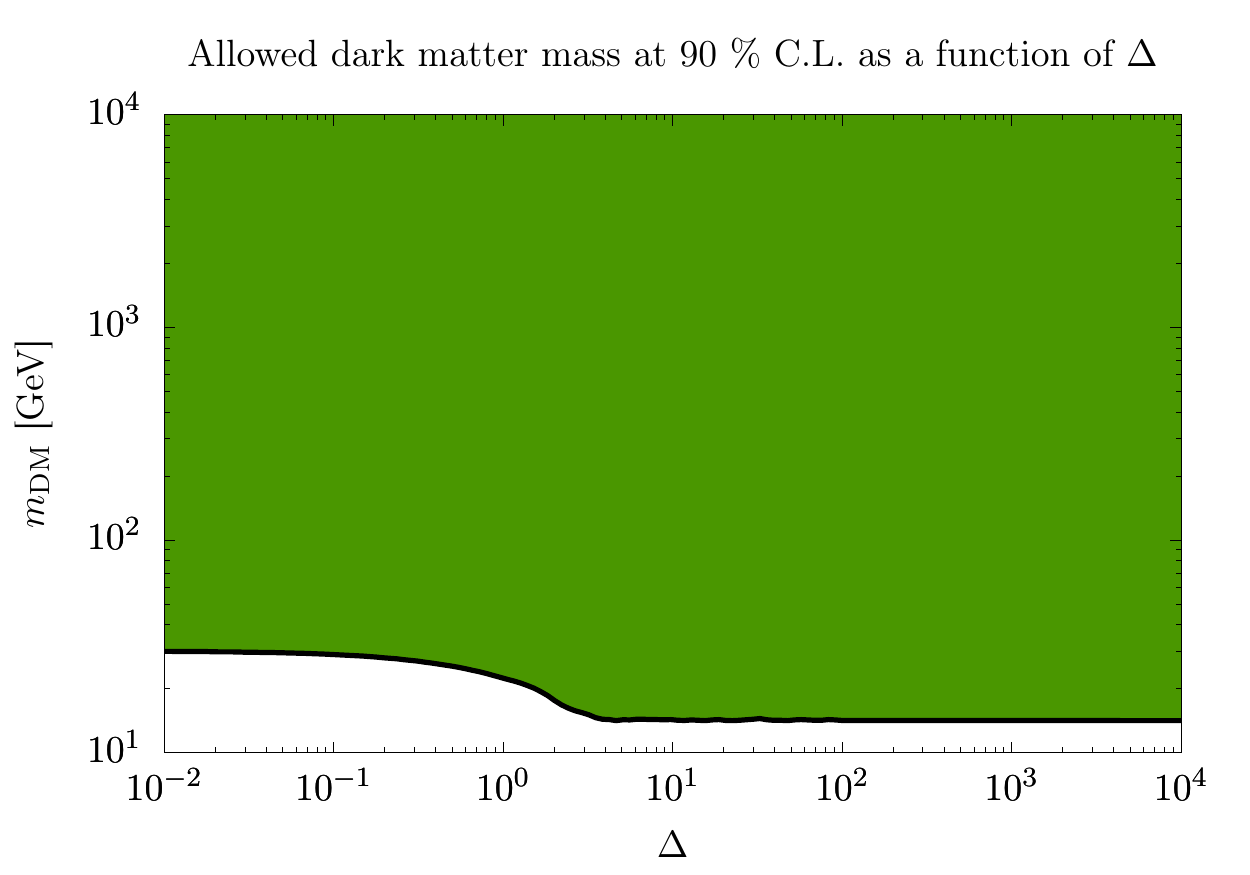}
	\end{center}
	\caption{Reconstruction of a putative signal after two years of XENON1T exposure from a WIMP with mass 50 GeV, spin-independent scattering cross section of $1.5 \times10^{-46}\,\text{cm}^2$ and SHM velocity distribution (marked by a white cross).
		The left plot shows the allowed region in the ($\mdm,\,\sigma_\text{SI}$) parameter space for selected values of $\Delta$, while the right plots depicts the range of allowed dark matter masses as a function of $\Delta$.}\label{fig:App3Reco}
\end{figure}

In Fig.~\ref{fig:App3Reco}, we show the projected 90\% CL allowed regions following a XENON1T discovery. Similar to what we saw in previous sections, the allowed regions cannot be extended to smaller values of $\sigma^p_\mathrm{SI}$ by more than a factor of $\sim 2$. For DM masses of $\sim 50\,\mathrm{GeV}$ and below, in the most agnostic case ($\Delta = 10^4$), the signal can be made consistent with very large cross sections. This is because XENON1T only probes the high speed tail of the velocity distribution in that case. A given cross section can then be made to fit the normalization of the data by reducing the fraction of the distribution above the threshold (see Ref.~\cite{Kavanagh:2013wba} for a related discussion). 

At intermediate masses, $100\,\mathrm{GeV} < m_\mathrm{DM} < 5000\,\mathrm{GeV}$, direct detection typically probes the majority of the velocity distribution, meaning that extreme values of $\Delta$ (that is, large distortions of the SHM) are required to hide the distribution function below the energy threshold of the experiment and reconcile the data with a large cross section. Indeed for deviations as large as $\Delta = 10^4$, the cross section is still constrained to within a factor of 10.

In the right panel of Fig.~\ref{fig:App3Reco}, we show the allowed ranges of the DM mass (profiled over $\sigma^p_\mathrm{SI}$) as a function of $\Delta$. We see that even without astrophysical uncertainties, the DM mass is only constrained to be greater than around $25 \,\mathrm{GeV}$. At large $\Delta$, this constraint weakens to $m_\mathrm{DM} \gtrsim 10\,\mathrm{GeV}$. In this case, the spectral shape of the signal is sufficiently different as to be irreconcilable with a $50\,\mathrm{GeV}$ WIMP, even with full freedom in the shape of $f(\vec{v})$.

\begin{figure}[t]
	\begin{center}
		\hspace{-0.75cm}
		\includegraphics[width=0.49\textwidth]{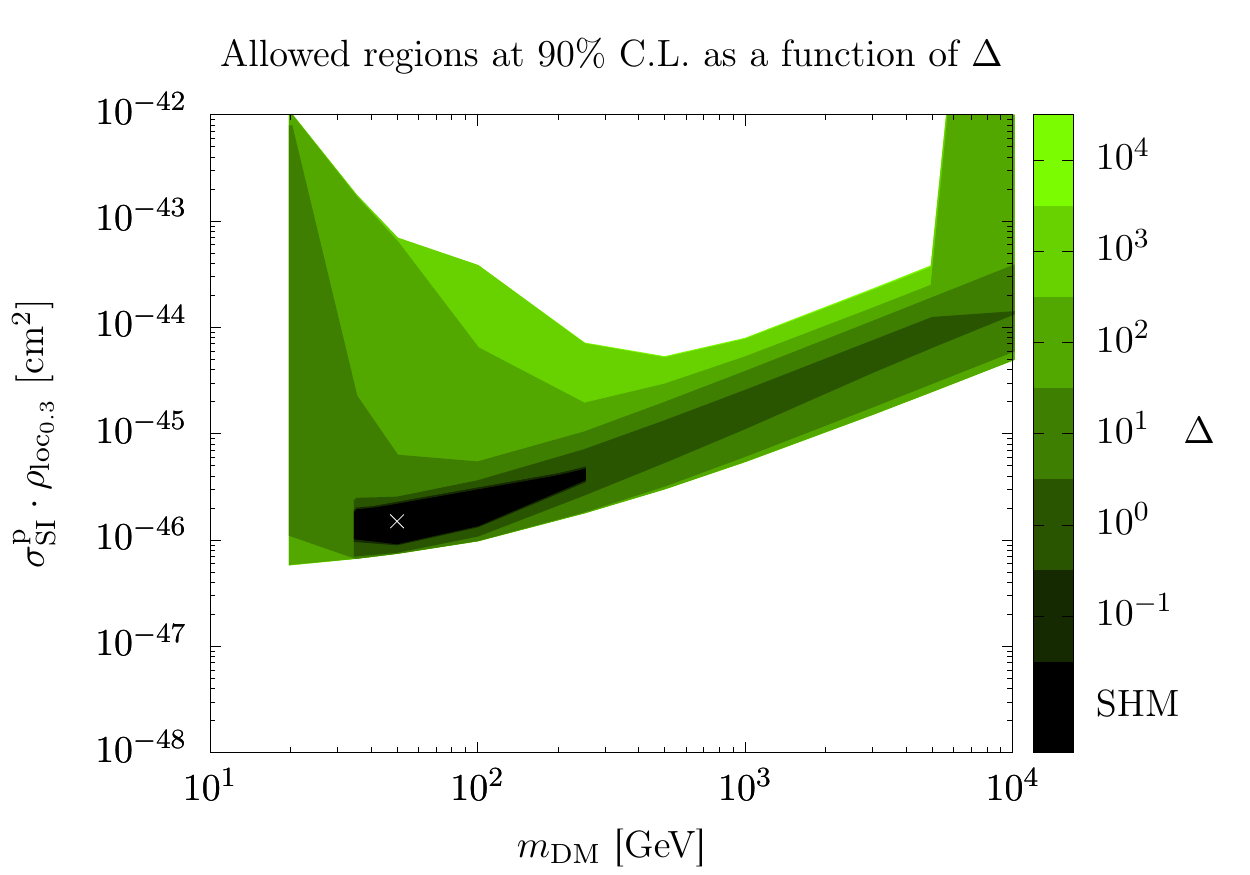}
		\includegraphics[width=0.49\textwidth]{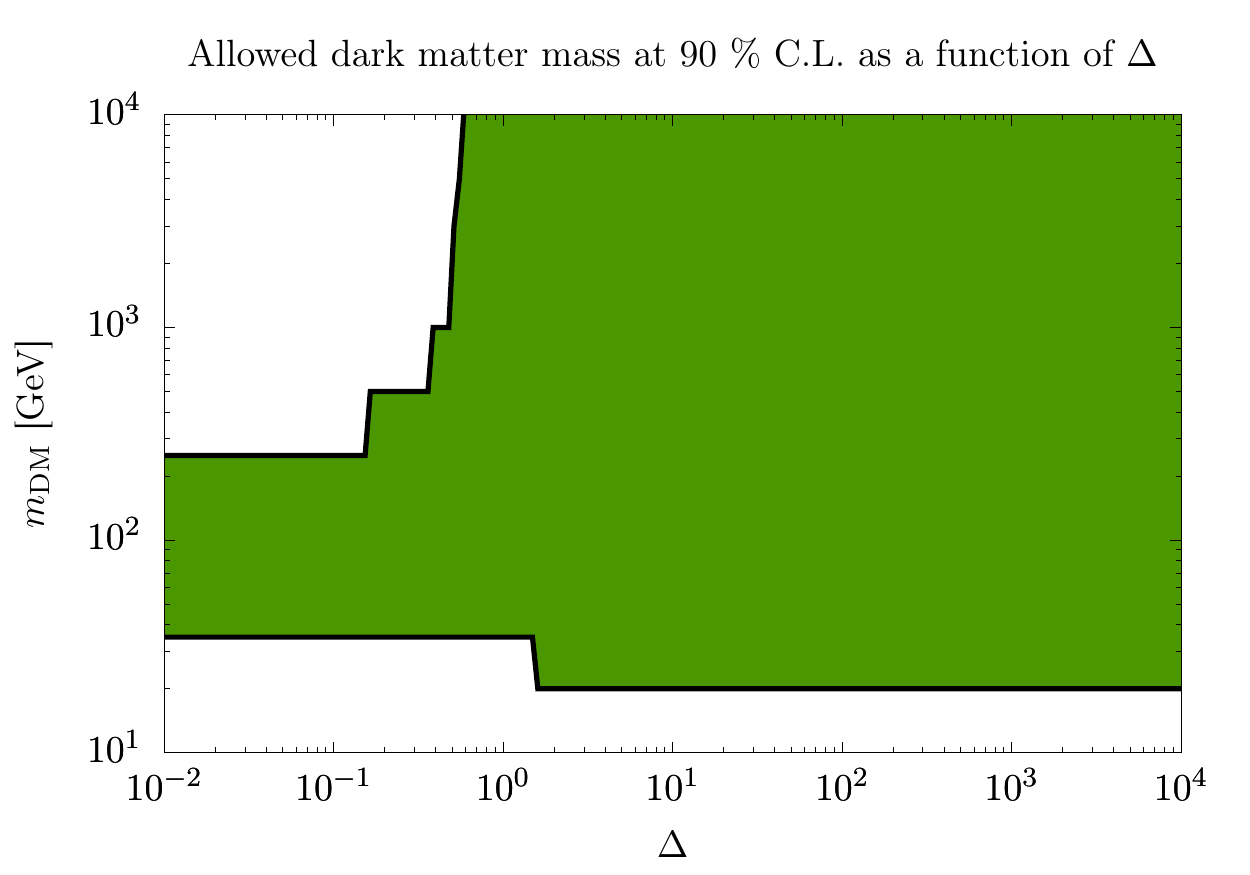}
	\end{center}
	\caption{Same as Fig.~\ref{fig:App3Reco} including also constraints from IceCube.}\label{fig:App3RecoIceCube}
\end{figure}

In Fig.~\ref{fig:App3RecoIceCube}, we show the same results (reconstructed contours and allowed masses), now obtained including the current constraints from IceCube. For IceCube, we use a single bin likelihood,
\begin{equation}
\label{eq:likelihood_IC}
\log\mathcal{L}_\mathrm{IC} = -\frac{1}{2} \frac{\left(N_\mathrm{sig} + N_\mathrm{BG} - N_\mathrm{obs}\right)^2}{\sigma^2}\,,
\end{equation}
where the number of observed events, background events and uncertainties are given in Eq.~\eqref{eq:ICdata}. The total log-likelihood is the sum of $\log\mathcal{L}_\mathrm{Xe}$ and $\log\mathcal{L}_\mathrm{IC}$. The current IceCube analysis is not sufficiently sensitive to detect our benchmark signal, assuming a Maxwell-Boltzmann velocity distribution. Nonetheless, it provides information which is useful in constraining the most extreme velocity distributions.

In Fig.~\ref{fig:App3RecoIceCube}, we see that the allowed regions in $\sigma^p_\mathrm{SI}$ are bounded from above for almost all values of $m_\mathrm{DM}$. For the XENON1T-only results, large cross sections could only be reconciled with the data by putting most of the distribution at low speed where it does not contribute to the rate. However, this low speed population would contribute to Solar capture and violate the bounds from IceCube. At intermediate masses, $\mathcal{O}(1\,\mathrm{TeV})$, this means that the cross section is constrained to within a factor of 10 for \textit{all} values of $\Delta$.

Concerning the reconstruction of the mass, it is clear that including astrophysical uncertainties widens the allowed values compared to the SHM, as the spectral information which is typically pins down the mass is degenerate with changes in the velocity distribution. However, we see in the right panel of Fig.~\ref{fig:App3RecoIceCube} that $\Delta \sim \mathcal{O}(1)$ distortions to the velocity distribution are required before this degeneracy begins to dominate and the allowed values of $m_\mathrm{DM}$ widen and extend to infinitely large mass. 

As described in Ref.~\cite{Kavanagh:2014rya}, a signal in a future neutrino telescope experiment (in addition to a XENON1T signal) would allow further improvements to the reconstruction. Here, we have improved on the method of Ref.~\cite{Kavanagh:2014rya} in that the optimization techniques we use here are substantially faster (computationally) than performing a full scan over velocity distribution parameters. While the analysis in Ref.~\cite{Kavanagh:2014rya} can be applied to an arbitrary likelihood, one could imagine using the optimization technique presented here to obtain an approximate map of the parameter space, refined later with a full likelihood scan, thus saving a great deal of computational effort.

\section{Conclusions}
\label{sec:Conclusions}

In this work, we have presented a technique for determining (for a fixed local DM density) the most aggressive and conservative limits in direct detection and neutrino telescope experiments, optimizing over arbitrary velocity distributions. Uncertainties in the local DM density can then be accounted for by a trivial rescaling of these limits. We also assign a statistically concrete meaning to combined limits and to possible future detections of dark matter. Being completely agnostic of the true form of the DM velocity distribution allows us to bracket the impact of astrophysical uncertainties as well as reduce bias in the reconstructed DM parameters from a putative signal. A similar bracketing of astrophysical uncertainties was presented in Ref.~\cite{Fairbairn:2012zs}, considering in particular non-Maxwellian velocity distributions. We go beyond that work by allowing deviations from the Maxwell-Boltzmann distribution of variable size, parameterized by $\Delta$, and ultimately by extending to the most general distributions possible ($\Delta \rightarrow \infty$). 

\begin{figure}[t!]
\hspace{-0.75cm}
	\begin{center}
		\includegraphics[width=0.49\textwidth]{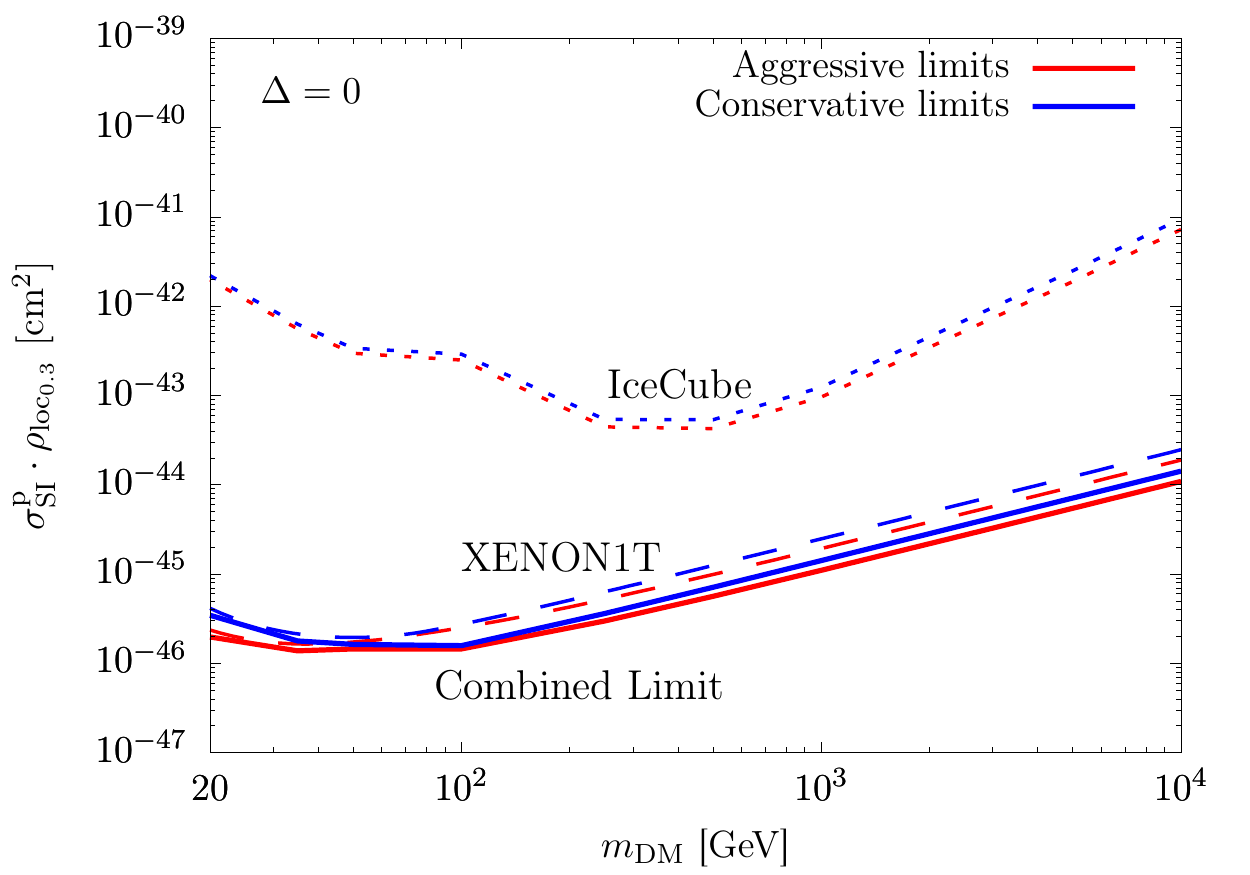}
		\includegraphics[width=0.49\textwidth]{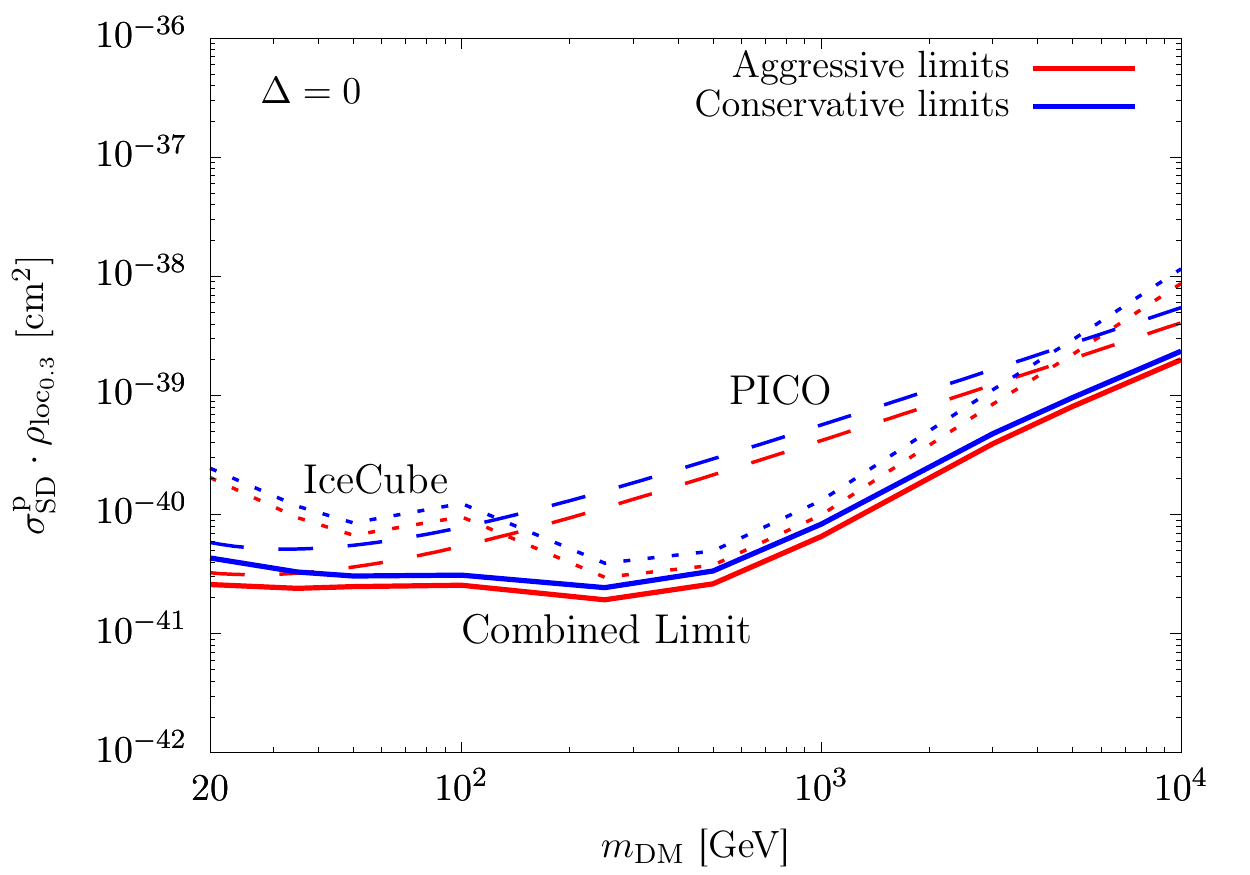}
		
		\includegraphics[width=0.49\textwidth]{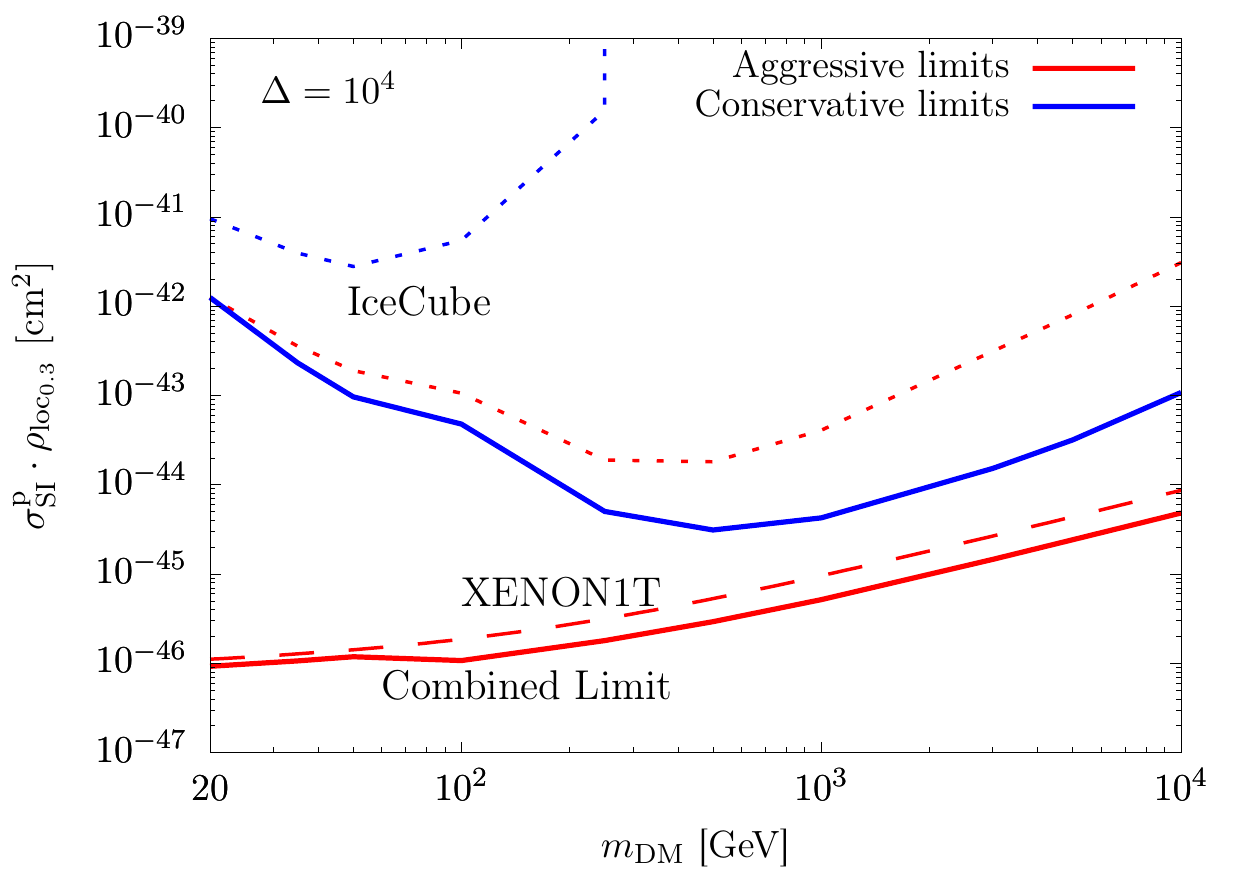}
		\includegraphics[width=0.49\textwidth]{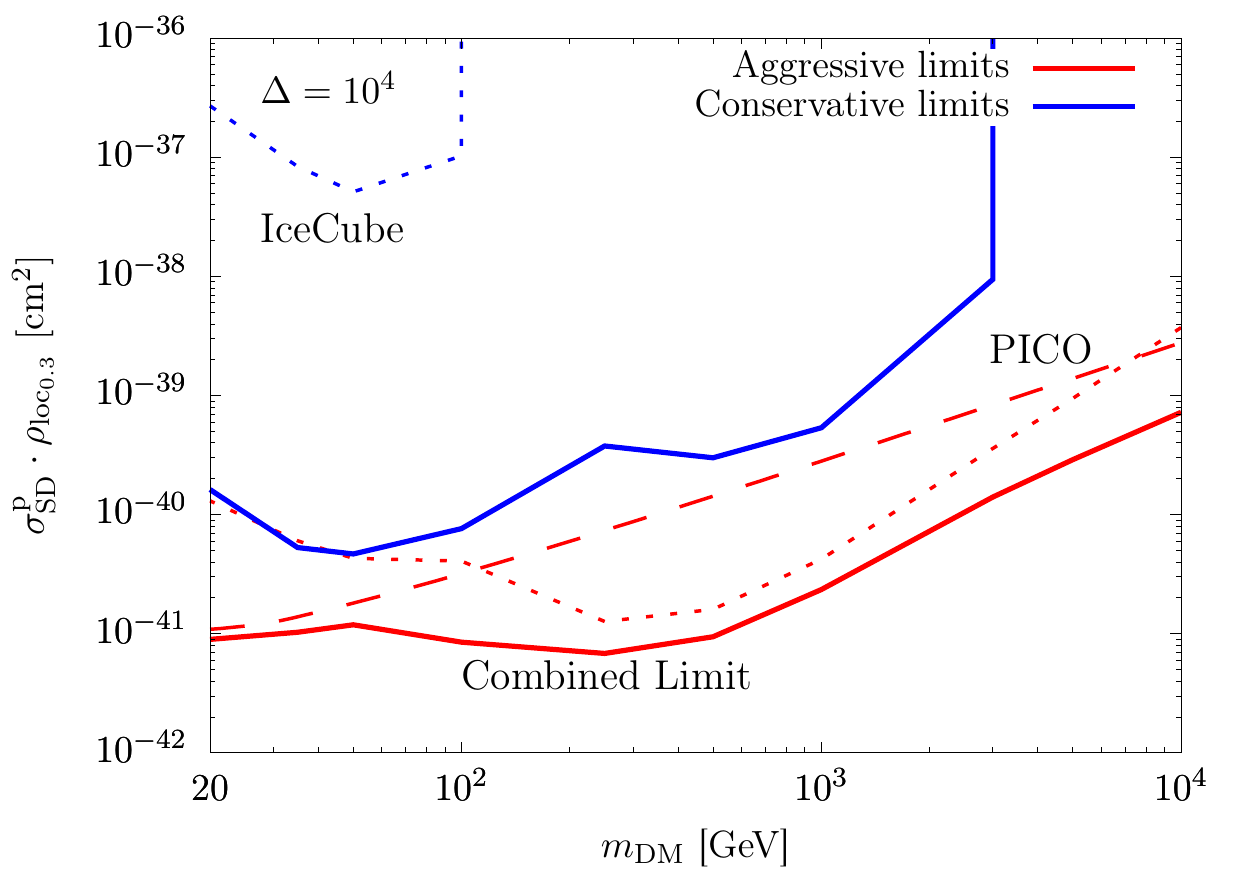}
	\end{center}
	\caption{Summary of the most aggressive (red) and most conservative (blue) upper limits on SI (left panels) and SD (right panels) interactions, obtained from direct detection and neutrino telescope data. In each panel, we show the limits from direct detection (XENON1T/PICO) alone (dashed), IceCube alone (dashed), and the combined limit (solid). The top row includes only uncertainties on the SHM ($\Delta = 0$) while the bottom row includes the possibility of arbitrary velocity distributions (allowing distortions up to $\Delta = 10^4$. Note that in some cases the most conservative limits are not shown, as they extend up to arbitrarily large cross sections.}\label{fig:aggressive_all}
\end{figure}

In the main body of the paper, we have applied our optimization techniques in a number of scenarios. In Fig.~\ref{fig:aggressive_all}, we summarise the most aggressive and most conservative upper limits which we have obtained. We show limits from considering direct detection and neutrino telescopes separately, as well as combined limits, for $\Delta = 0$ (SHM) and $\Delta = 10^4$ (arbitrary velocity distributions). We summarize our key conclusions below:
\begin{itemize}
\item Allowing for arbitrary velocity distributions, DM direct detection limits can be strengthened by at most a factor of 2. In contrast, direct detection limits may be weakened almost indefinitely, by populating the DM distribution only below the energy threshold of the experiment.
\item Neutrino telescope limits can be strengthened by at most a factor of 2.5. For large DM masses, the most conservative limits vanish while for small masses (for which capture is more kinematically favourable) the conservative limit is finite and a factor of $10$-$10^3$ times weaker than the SHM. This limit cannot be evaded or weakened, as long as we assume that there is a population of DM gravitational bound to the Milky Way.
\item Combining neutrino telescope and direct detection limits (in a statistically consistent way), we find that for almost all DM masses, the most conservative bounds (allowing full freedom in $f(\vec{v})$) remain finite. For masses in the range 100-1000 GeV, the most aggressive and most conservative combined constraints differ only by an order of magnitude. In this mass range, the constraints are relatively robust to astrophysical uncertainties.
\item In the event of a discovery in direct detection experiments, including astrophysical uncertainties greatly extends the range of masses and cross sections consistent with the data. However, including data from a neutrino telescope typically constrains the interaction cross section from above. In addition, the mass of the DM particle can still be well-constrained even allowing up to $\mathcal{O}(1)$ deviations from the standard Maxwell-Boltzmann distribution.
\end{itemize}

The approach we present -- decomposing the velocity distribution as a large number of streams and optimizing over the stream densities -- is general, requiring only minimal assumptions, and can be applied to arbitrary likelihoods and analyses. Typically, this optimization problem involves a very large number of free parameters and could be computationally impractical using standard techniques such as Monte Carlo parameter scans. Instead, we have made use of optimization techniques from quadratic programming, appropriate for calculating event numbers and test-statistics which are at most quadratic in the DM velocity distribution $f(\vec{v})$. As we have demonstrated, this is approximately the case for some rare event searches and, in the event of a discovery, will become a better and better approximation as more data is acquired and the Gaussian limit is approached. However, the limits we present remain approximate, neglecting also systematic and background uncertainties in the experiments we consider. 

Even so, our approach allows us to map out the approximate range of allowed DM parameters in a fast, tuneable and general way. In the event of a discovery, for example, this map could then be refined with exact but computationally more expensive methods (such as that of Ref.~\cite{Kavanagh:2014rya}) to obtain general astrophysics-independent results. Furthermore, our approach allows us to understand how much astrophysical uncertainties could affect conclusions from DM searches in the most extreme cases. As we have seen, limits from direct detection experiments are only a factor of a few weaker than the most aggressive possible limits. Furthermore, combined limits from direct detection experiments and neutrino telescopes cannot be weakened indefinitely and in many cases the most conservative limits are with a factor of 10 of the most aggressive ones. This highlights that -- as long as astrophysical uncertainties are properly accounted for -- such experiments remain a robust and powerful probe of  dark matter in the Solar System. 

\section*{Acknowledgments}
This work has been supported by the DFG cluster of excellence EXC 153 ``Origin and Structure of the Universe'' and by the Collaborative Research Center SFB1258. BJK gratefully acknowledges support from the Netherlands Organisation for Scientific Research (NWO) through the VIDI research program ``Probing the Genesis of Dark Matter" (680-47-532). We would like to thank the Munich Institute for Astro- and Particle Physics (MIAPP) where part of this work was developed, and especially to Sebastian Wild for useful discussions on the XENON1T limits. We would also like to thank the `anonymous' referee for their helpful remarks.

\bibliographystyle{JHEP-mod}
\bibliography{references}
\end{document}